\pdfoutput=1
\PassOptionsToPackage{x11names,svgnames}{xcolor}
\PassOptionsToPackage{left=1.5cm,right=1.5cm}{geometry}
\documentclass[submission, Phys]{SciPost}

\hypersetup{
	unicode=true,
	pdftoolbar=true,
	pdfmenubar=true,
	pdffitwindow=false,
	pdfstartview={FitH},
	pdftitle={Strength in numbers: optimal and scalable combination of LHC new-physics searches},
	pdfauthor={Jack~Y.~Araz,Andy~Buckley,Benjamin~Fuks,Humberto~Reyes-Gonzalez,Wolfgang~Waltenberger,Sophie~L.~Williamson,Jamie~Yellen},
	pdfnewwindow=true,
	colorlinks=true,
	linkcolor=blue,
	citecolor=magenta,
	filecolor=magenta,
	urlcolor=cyan
}

\usepackage{xspace}
\usepackage{afterpage}
\usepackage{graphicx}
\usepackage{relsize}
\usepackage{mdframed}
\usepackage[toc,page]{appendix}
\usepackage{fancyvrb}
\usepackage{placeins}
\usepackage{longtable,multirow}
\usepackage{hyperref}
\usepackage{url}
\usepackage{subfig}
\usepackage{blindtext}
\usepackage{enumitem}
\usepackage{lineno}

\usepackage{amssymb}
\usepackage{amsfonts}
\usepackage{amsmath}
\usepackage{subdepth}

\usepackage{algorithm,algpseudocode}

\usepackage{todonotes}
\usepackage{xpatch}
\makeatletter
\xpretocmd{\todo}{\@bsphack}{}{}
\xapptocmd{\todo}{\@esphack}{}{}
\makeatother

\newcommand{\slw}[1]{\slw[]{#1}}

\newcommand{\sw}[1]{\textsc{#1}\xspace}

\newcommand{\ie}{\emph{i.e.}}

\newcommand{\amc}{\textsc{MadGraph5\textunderscore}a\textsc{MC@NLO}\xspace}

\newcommand{\pythiaE}{\textsc{Pythia\,8}\xspace}

\newcommand{\delphes}{\textsc{Delphes\,3}\xspace}
\newcommand{\fastjet}{\textsc{FastJet}\xspace}
\newcommand{\lhapdf}{\textsc{Lhapdf}\xspace}
\newcommand{\MA}{\textsc{MadAnalysis\,5}\xspace}
\newcommand{\CM}{\textsc{CheckMATE}\xspace}
\newcommand\smodels{\textsc{SModelS}\xspace}
\newcommand\rivet{\textsc{Rivet}\xspace}
\newcommand\contur{\textsc{Contur}\xspace}
\newcommand\gambit{\textsc{Gambit}\xspace}
\newcommand{\taco}{\textsc{taco}\xspace}

\newcommand{\SR}{\text{SR}\xspace}
\newcommand{\SRs}{\text{SRs}\xspace}

\usepackage{siunitx}
\DeclareSIUnit{\electronvolt}{\text{e\kern-0.15ex V}}
\DeclareSIUnit{\eV}{\electronvolt}
\DeclareSIUnit{\MeV}{\mega\eV}
\DeclareSIUnit{\GeV}{\giga\eV}
\DeclareSIUnit{\TeV}{\tera\eV}
\DeclareSIUnit{\ifb}{\ensuremath{\text{fb}^{-1}}}

\usepackage{microtype}
\makeatletter
\g@addto@macro\bfseries{\boldmath}
\let\@to\to
\renewcommand{\to}{\ensuremath{\@to}}

\makeatother

\usepackage{fixltx2e,relsize}
\newcommand{\xtitle}[1]{\def\thetitle{#1}}
\newcommand{\xauthor}[2][]{{#2}\textsuperscript{#1}}
\newcommand{\xaddress}[2][]{\textsmaller{\textbf{#1}~\textit{#2}}\newline}

\begin{document}

\xtitle{Strength in numbers: optimal and scalable combination of LHC new-physics searches}
\begin{center}
  \LARGE \bfseries \thetitle
\end{center}

\medskip

\begin{center}
\xauthor[a]{Jack~Y.~Araz},
\xauthor[b]{Andy~Buckley},
\xauthor[c]{Benjamin~Fuks},
\xauthor[d,e]{Humberto~Reyes-Gonzalez}, 
\xauthor[f,g]{Wolfgang~Waltenberger},
\xauthor[h]{Sophie~L.~Williamson},
\xauthor[b]{Jamie~Yellen}
\end{center}

\begin{center}
\begin{minipage}{0.9\textwidth}
\xaddress[a]{Institute for Particle Physics Phenomenology, Durham University, Durham, UK}
\xaddress[b]{School of Physics and Astronomy, University of Glasgow, Glasgow, United Kingdom}
\xaddress[c]{Laboratoire de Physique Th\'eorique et Hautes \'Energies (LPTHE), UMR 7589, Sorbonne Universit\'e et CNRS, 4 place Jussieu, 75252 Paris Cedex 05, France}
\xaddress[d]{Department of Physics, University of Genova, Via Dodecaneso 33, 16146 Genova, Italy}
\xaddress[e]{INFN, Sezione di Genova, Via Dodenasco 33, I-16146 Genova, Italy}
\xaddress[f]{Institut f\"ur Hochenergiephysik, \"Osterreichische Akademie der Wissenschaften, Nikolsdorfer Gasse 18, 1050 Wien, Austria}
\xaddress[g]{University of Vienna, Faculty of Physics, Boltzmanngasse 5, A-1090 Wien, Austria}
\xaddress[h]{Institute for Theoretical Physics, Karlsruhe Institute of Technology,\\ 76128 Karlsruhe, Germany}
\end{minipage}
\end{center}

\begin{abstract}
To gain a comprehensive view of what the LHC tells us about physics beyond the
Standard Model (BSM), it is crucial that different BSM-sensitive analyses can be
combined. But in general search-analyses are not
statistically orthogonal, so performing comprehensive combinations requires
knowledge of the extent to which the same events co-populate multiple analyses'
signal regions. We present a novel, stochastic method to determine this degree
of overlap, and a graph algorithm to efficiently find the combination of
signal regions with no mutual overlap that optimises expected upper limits on
BSM-model cross-sections. The gain in exclusion power relative to single-analysis
limits is demonstrated with models with varying degrees of complexity, ranging
from simplified models to a 19-dimensional supersymmetric model.
\end{abstract}




\clearpage
\tableofcontents

\section{Introduction}\label{sec:intro}


The ATLAS and CMS experiments at the Large Hadron Collider (LHC) are performing
direct searches for new physics beyond the Standard Model (BSM) in many
different channels. The previous decade of LHC operation has already put strong
constaints on the most obvious models of BSM physics, pushing their viable
configurations to arguably untenable high new-particle masses; compared to these
simple models pushed to extreme configurations, it is natural also for models
with subtler phenomenology to enter the new-physics discourse. Such models bring
increasing complexity in both the dimensionality of their parameter spaces, and
the range of phenomenology possible within them. This leads to an increasing
presumption that new physics will not be discoverable via a single, powerful
experimental signature, but will disperse across many signatures at a level
below direct exclusion in any one search analysis.

These factors cause a major logistical headache for LHC data-interpretation. At
the fully detector-simulated level used for experiment interpretations, adaptive
samplings and scans of high-dimensional spaces are not feasible, yet the few-parameter
simplified-model approach used in the first LHC runs will no longer be
representative of the analysis power to constrain actually viable BSM models.
In this mode, it is clear that analyses must be systematically combined
together, and initial scoping of viable parameter-space regions performed via a
more lightweight approximation of experiment response, such as via the
\MA~\cite{Conte:2018vmg}, \CM~\cite{Drees:2013wra},
\rivet+~\contur~\cite{Bierlich:2019rhm,Buckley:2021neu},
\gambit~\cite{GAMBIT:2017yxo}, or similar toolkits.

In this paper we focus on the first problem: how to best combine analyses for
optimal statistical significance, which for the purposes of our analysis is
the ability to exclude a specific BSM model point at fixed confidence level.
 Definitive LHC statements about any dispersed signature will require
combination of as many analyses as possible, but not all analyses \emph{can} be
combined. Were we simply to combine the test statistics of every signal region
(SR) from every analysis available in the public collections, we would certainly
double-count physics effects, since the same events will manage to pass multiple
analyses' event-selection cuts and observable binnings.

Different sets of observables are used for selection-cut purposes in each
analysis, but the disjoint choices are typically highly correlated through a
complex dependence on the rest of the selection phase-space. It is hence impossible to
reliably identify degrees of overlap directly from a list of cut observables and
values. And even when the analysis overlaps \emph{are} known, there remains the
problem of identifying which compatible subset will place the optimal constraints
on any given BSM model.

Approaches so far have hence been manual, and rather
conservative~\cite{Buckley:2021neu,protomodels}. To scale up to the full set of
LHC legacy analyses at 7, 13, and \SI{13.6}{\TeV}, and to obtain maximal
statistical limits from the resulting combination, a more quantitative and
automated approach is needed. This paper provides a blueprint for such an
approach: we use the \MA analysis toolkit in conjunction with
\smodels~\cite{Kraml:2013mwa} to estimate degrees of analysis overlap over
hundreds of signal regions, and propose a new graph-based algorithm
to optimise the subset of non-overlapping analyses used for testing a given
BSM model.

\bigskip

In this work we determine the parameter space accessed by the topologies which
populate the different signal regions contained in a sample of 18 CMS and ATLAS
analyses, the current maximum overlap between the \MA and \smodels
re-interpretation toolkits. While far from a complete set, this is sufficient to
illustrate the complexity of (undeclared) overlapping SR acceptances, and the
non-triviality of identifying the most significantly exclusionary, combinable
subset of SRs for a given BSM model.

In Section~\ref{sec:correst}, by sampling over the minimal parameter spaces of
the \smodels simplified-model topologies able to populate these SRs, and using a
version of \MA modified to provide information for per-event Poisson bootstrapping, we
estimate the statistical overlaps between the resulting set of 355~signal
regions. Applying a threshold on the degree of overlap acceptable in combination then
results in a matrix of acceptable SR--SR combinations, from which the space of
optimal subsets can be explored.

In Section~\ref{sec:srcomb}, we find that a powerful method for doing so is to
represent the SRs in a graph-theoretic form, in which sensitivity maximisation
for a variety of physics-performance metrics can be formalised as a weighted
longest-path problem.

In Section~\ref{sec:results}, we apply this technique on a series of
increasingly complex and general BSM models, ranging from ``closure tests'' on
simplified models, to compact models with dispersed phenomenology, to the
19-dimensional phenomenological Minimal Supersymmetric Standard Model (pMSSM-19).

We conclude with reflections on what is needed in technical and
community-coordination terms to bring this method and the resulting gains in LHC
physics sensitivity to practical realisation.
\section{Overlap estimation}
\label{sec:correst}

To investigate how analyses can be combined to provide the most stringent
constraints on a BSM model point, we choose the selection of analyses available
both in \smodels and \MA as our database. At the time
of writing this includes 18 analyses:
ATLAS-SUSY-2013-02~\cite{ATLAS:2014jxt}, %
ATLAS-SUSY-2013-04~\cite{ATLAS:2013qzt}, %
ATLAS-SUSY-2013-05~\cite{ATLAS:2013lcn}, %
ATLAS-SUSY-2013-11~\cite{ATLAS:2014zve}, %
ATLAS-SUSY-2013-21~\cite{ATLAS:2014hqe}, %
ATLAS-SUSY-2015-06~\cite{ATLAS:2016dwk}, %
ATLAS-SUSY-2016-07~\cite{ATLAS:2017mjy}, %
ATLAS-SUSY-2018-04~\cite{ATLAS:2019gti}, %
ATLAS-SUSY-2018-06~\cite{ATLAS:2019wgx}, %
ATLAS-SUSY-2018-31~\cite{ATLAS:2019gdh}, %
ATLAS-SUSY-2018-32~\cite{ATLAS:2019lff}, %
ATLAS-SUSY-2019-08~\cite{ATLAS:2020pgy};
CMS-SUS-13-011~\cite{CMS:2019vzo}, %
CMS-SUS-13-012~\cite{CMS:2014tzs}, %
CMS-SUS-16-033~\cite{CMS:2017abv}, %
CMS-SUS-16-039~\cite{CMS:2017moi}, %
CMS-SUS-16-048~\cite{CMS:2018kag}. %
CMS-SUS-17-001~\cite{CMS:2017jrd}, and
CMS-SUS-19-006~\cite{CMS:2019zmd}.

The cascade decays, or \emph{topologies}, covered by these analyses are
simplified so that they focus on the production of two massive BSM states that
each decay to at most 2--3 final-state particles. 
The topologies covered by these analyses, using the \smodels naming
convention~\cite{smodels:dictionary}, are:
T1, T1bbbb, T1btbt, T1tttt(-off), T2, T2bb, T2tt(-off), T2bbWW(-off), T2bt,
T2cc, T3GQ, T5, T5bbbb, T5tctc, T5tttt, T5GQ, T5WW(-off), T5WZh, T5ZZ, T6bbhh,
T6bbWW(-off), T6WW(-off), T6WZh, TChiChipmSlepL, TChiChipmSlepStau,
TChiChipmStauStau, TChiChipmSlepSlep, TChipChimSlepSnu, TSlepSlep, TChiZZ
TChiWH, TChiWW, TChiWZ(-off), TChiZoff, TGQ, TSlepSlep, and TStauStau.

\subsection{Model-space sampling and event generation}
\label{sec:sampling}

In order to obtain robust conclusions about potential signal overlaps for arbitrary scenarios, we proceed as follows. For each analysis, we construct a convex hull in each simplified model's parameter space that is accessed by a given topology, carried out using the efficiency maps implemented in \smodels~\cite{smodels:listofanalyses}. The efficiency maps give upper limits on the production cross-sections of the two relevant BSM states, and depend on the masses in the simplified decay chains. For each simplified model, one such convex hull exists for each analysis that has a result for that given simplified model. We are interested in the joint set of convex hulls corresponding to each simplified model. Thus, we construct  a contour around the mass-parameter space beyond which the expected event-yield from all corresponding analyses is zero. In this way, the union of regions will be populated with events, without multiply populating those shared between analyses. We uniformly generate events within this joint convex hulls, so to only introduce an uninformative flat prior in our procedure.


The MC events were generated at LO with \amc~v2.6.5~\cite{Alwall:2014bza} at the
partonic level with the NNPDF~2.3~LO~\cite{Ball:2013hta} set of parton
distribution functions via the \lhapdf library~\cite{Buckley:2014ana}, with
parton-showering and hadronisation simulated by \pythiaE~\cite{Sjostrand:2007gs}
through the \amc interface. Detector-level events were obtained with \delphes
and \fastjet~\cite{deFavereau:2013fsa,Cacciari:2011ma}, executed through \MA
with analysis-specific configurations interleaved with the event-selection
logic. The input for the generation pipeline was a corresponding
SLHA-format~\cite{Allanach:2008qq} data-file for each topology, with the masses
of the produced, final, and (in some cases) intermediate BSM states defined as
free parameters. The initial partonic processes in the generation chain were in
all cases direct production of the topology's massive BSM states, with decay
chains implemented via \pythiaE's decay mechanism.

The required output of the MC generation procedure is a binary \emph{acceptance
matrix} $\Theta$ of shape $N_\mathrm{evt} \times N_{\SR}$, where $\Theta_{e,s} =
1$ means that event $e$ populated SR-bin $s$, and \emph{vice versa}
$\Theta_{e,s} = 0$ when event $e$ did not pass the cuts for SR $s$.\footnote{In
general this matrix need not be binary, and can represent a per-event bin yield,
for observables that can have multiple fills per event. Event weighting also
complicates matters. But in the current context of binary acceptance or
rejection of unweighted events, $\Theta_{ij} \in \{0,1\}$.}
This matrix is produced using the new
\begin{verbatim}
set main.recast.TACO_output = <file-name>
\end{verbatim}
command in \MA, added to the framework for this purpose. The \emph{acceptance
matrices} are emitted as text files with each event corresponding to a pair of
lines encoding first the list of floating-point event
weights~\cite{Bothmann:2022pwf} (in this study we use only the nominal weight),
and then a list of \texttt{0} and \texttt{1} characters corresponding to the
$N_\SR$ signal regions. These files are written separately for each \delphes
configuration to the location
\begin{verbatim}
<Output>/SAF/defaultset/<delphes-card-name>.<file-name>
\end{verbatim}
in the output directory of the recasting process.

To determine the minimal number of Monte-Carlo events needed for a reliable
estimation of the overlap matrix, we start with 100 events for \num{1000}
random parameter points sampled from the union of the convex hulls of the
signal regions, so with an initial N=\SI{100000}\, events.
For any pair of signal regions $\SR_1$ and $\SR_2$
populated with $n_1$ and $n_2$ events respectively, we then determine the number
$k$ of shared events. If $k>100$, we have accumulated enough statistics, and
proceed to the bootstrapping procedure.
For $k \le 100$, with $n \equiv n_1 + n_2$, we use
the confidence interval construction by Clopper and Pearson~\cite{ClopperPearson34} of the
binomial distribution 
\begin{equation}
\mathrm{B}(n,p)=\binom{n}{k}p^k(1-p)^{n-k} 
\end{equation}
where $p$ is a free parameter defined as the \emph{probability of overlap}. In
order to guarantee enough events for the case of a negligible overlap, we
need to obtain a one-sided (upper) confidence interval for $p$ at confidence level
$\mathrm{CL} \equiv 1-\alpha = 0.95$ and guarantee that it is below a
certain threshold.
From the Clopper--Pearson construction, this is computed as the $1-\alpha$
quantile of the $\beta$ distribution
\begin{equation}
  f(p;k,n) = \beta_{1-\alpha;k+1;n-k}.
\end{equation}

If this upper bound is below the arbitrarily chosen threshold, $f<0.01$, we assume that we have accumulated
enough statistics to safely infer the potential absence of a significant overlap, and we
confidently proceed to the bootstrapping procedure.

\begin{figure}
  \centering
  \includegraphics[width=0.60\textwidth]{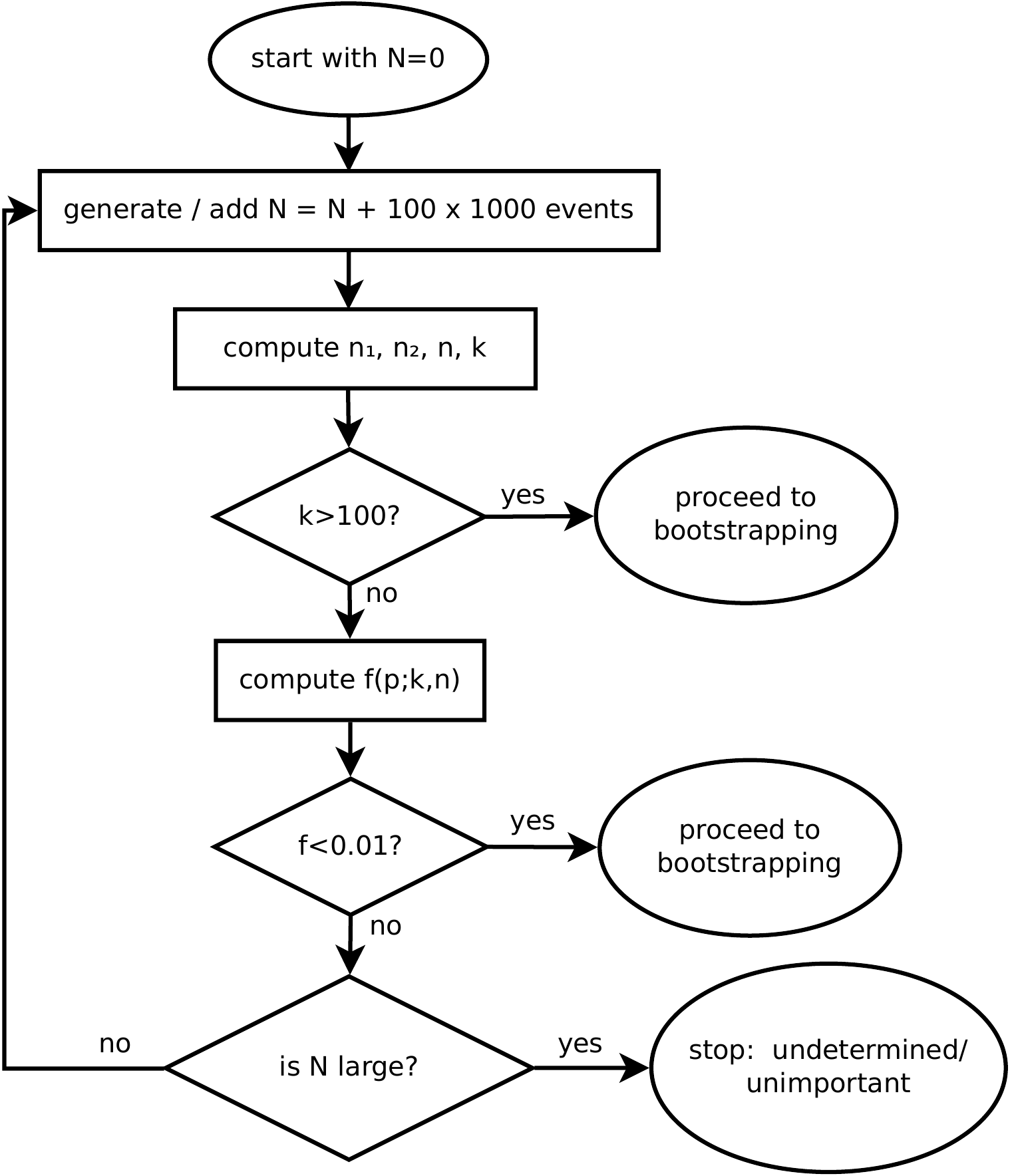}
  \caption{Flowchart for the determination of the number of Monte Carlo events
    needed for estimation of the overlap matrix.}
  \label{fig:clopper_pearson}
\end{figure}

Using these criteria we can employ the logic of
Figure~\ref{fig:clopper_pearson}. It will guarantee that enough statistics is
available to robustly and reliably determine both a significant or negligible overlap between a given pair of signal regions.

\subsection{Overlap-matrix estimation}
\label{sec:corrindep}

Once with a set of sufficiently populated SRs, we are ready to determine whether
or not such SRs are approximately orthogonal with respect to one another.

The Pearson correlation can be estimated from the acceptance matrix via the
event-averaged acceptance covariance,
\begin{equation}
  \begin{split}
    \mathrm{cov}_{ij}
    &= \langle \Theta_i \Theta_j \rangle - \langle \Theta_i \rangle \langle \Theta_j \rangle\\
    &\equiv \frac{\sum_e \Theta_{e,i} \Theta_{e,j}}{N_\mathrm{evt}}  - \frac{\sum_{e'} Y_{e'\!,\,i} ~\cdot~ \sum_{e''} Y_{e''\!\!,\,j}}{N_\mathrm{evt}^2}  \, ,
  \end{split}
\end{equation}
where $N_\mathrm{evt}$ is the number of events in the estimation sample and,
as made explicit in the second line, $i$ and $j$ are SR indices. This method
is possible because the entire event-wise acceptance matrix is available and hence
overlaps can be estimated by averaging over the event axis of the matrix.

An equivalent approach, taken by the current code, is to perform bootstrap
sampling from a unit Poisson distribution. Each event is assigned
$N_\mathrm{boot}$ random ``bootstrap weights'' $w_{e,b} \sim
\mathrm{Pois}(\lambda=1)$, which are aggregated on to $N_\mathrm{boot}$ replicas
of the SRs' yield estimates. The result is a $N_{\SR} \times N_\mathrm{boot}$
bootstrapped \emph{yield matrix} $Y$, which expresses the sum of event weights
falling into the set of SRs for each of the $N_\mathrm{boot}$ alternative
histories generated from the single set of input events. The overlaps between
SRs can then be determined from their common weight-fluctuations over the set
of histories, i.e.~another estimate of the covariance:
%
\begin{equation}
\begin{split}
  \mathrm{cov}_{ij}
  &= \langle Y_i Y_j \rangle - \langle Y_i \rangle \langle Y_j \rangle\\
  &\equiv \frac{\sum_b Y_{i,b} Y_{j,b}}{N_\mathrm{boot}}  - \frac{\sum_{b'} Y_{i,b'} ~\cdot~ \sum_{b''} Y_{j,b''}}{N_\mathrm{boot}^2}  \, .
\end{split}
\end{equation}
%
The distinction is that the averaging is now over bootstrap replicas of the
aggregate yields, rather than the per-event acceptance tuple.  While not
essential in the current implementation, the bootstrap approach avoids the need
to manage a linearly growing acceptance matrix, in favour of a fixed-size
$N_\mathrm{SR} \times N_\mathrm{boot}$ yield matrix, which may become computationally
relevant for large event samples.

From the covariance matrix, obtained through either strategy, we define the
\emph{overlap matrix}
\begin{equation}
  \rho_{ij} = \frac{ \mathrm{cov}_{ij} }{ \sqrt{\mathrm{cov}_{ii} \, \mathrm{cov}_{jj}} } \, ,
\end{equation}
following the usual Pearson-correlation definition. Lower-triangle plots of this
symmetric overlap matrix for the sets of signal regions common to \smodels and
\MA are shown in the appendix Figures~\ref{fig:overlaps8}
and~\ref{fig:overlaps13} for \SI{8}{\TeV} and~\SI{13}{\TeV} LHC data-analyses
respectively, with patterns of highly and partially co-populated SRs clearly
visible.

Finally, a binary \emph{exclusivity matrix} $E$ between SR-pairs $\SR_{i}$ and
$\SR_{j}$ is derived by applying an ``acceptable overlap'' threshold $T$ such
that the exclusivity between SRs $i$ and $j$ is $E_{ij} = (\vert \rho_{ij} \vert
\le T)$.  The value chosen for $T$ is at present somewhat subjective, reflecting
that for each use-case there will be a finite value of $\rho_{ij}$ below which
double-counting biases are not statistically resolvable: treating these low
correlations as zero-correlations avoids blocking useful SR combinations due to
irrelevant and noisy correlation estimates.

The procedure described above is implemented in the public \textsc{Python}
program TACO (Testing Analyses COrrelations), available
at~\url{https://gitlab.com/t-a-c-o/taco_code}.

\section{Optimal signal-region combination}
\label{sec:ranking}
\label{sec:srcomb}

Armed with the exclusivity matrix from the previous section, for a choice of
overlap threshold $T$, we now have the challenge of identifying the
best-expected combination of SRs compatible with it. We consider this in two
steps: first the combinatoric problem of efficiently constructing \emph{all}
allowed paths, and then the optimisations to this enabled by the specific
definition of ``best'' used in BSM-analysis reinterpretation.

\subsection{Compatible signal-region sets as path-finding}
\label{sec:pathfinding}

Without prior information about overlaps or statistical significances, the
process of finding a preferred subset of signal regions from a set of size $n$
is a combinatorial challenge with $2^n$ possible solutions.
Exhaustively generating and evaluating each such combination hence suffers from
exponential time-complexity scaling, and is computationally impractical even for
relatively small $n$, let alone the $n \sim \mathcal{O}(1000)$ required by real
reinterpretations.

However, considering the SR-acceptance exclusivities $E_{ij}$, the majority of
these $N$ combinations transpire to be \emph{forbidden}, as large-$r$ na\"ive
subsets become overwhelmingly likely to contain at least one overlapping pair of
SRs. The question then becomes whether, given prior awareness of $E_{ij}$, it is
possible to evaluate all \emph{allowed} SR-combinations more efficiently than
exhaustive generation followed by overlap-checking.


In this section we show that the answer to this question is yes, and that the
problem can be usefully recast as finding an optimum path through a directed
acyclic graph (DAG). We present an algorithm that reduces the asymptotic time
complexity by efficiently selecting path-elements based on recursive application
of the SR exclusivity matrix, rendering the combinatoric problem not just
tractable but computationally fast.

The key insight is to avoid generating invalid SR-combinations at all: this can
be achieved by generating the combinations directly from the overlap matrix.
Hence we must restrict the generated subsets only to those for which $E_{ij} =
1$ for all distinct $i,j$ in the set. This condition requires that if a subset
of all possible signal regions is built up iteratively, its $j${th} element
must have no significant overlap with all the previously selected elements
$0 ... j-1$.

\begin{figure}
  \centering
  \includegraphics[width=0.60\textwidth]{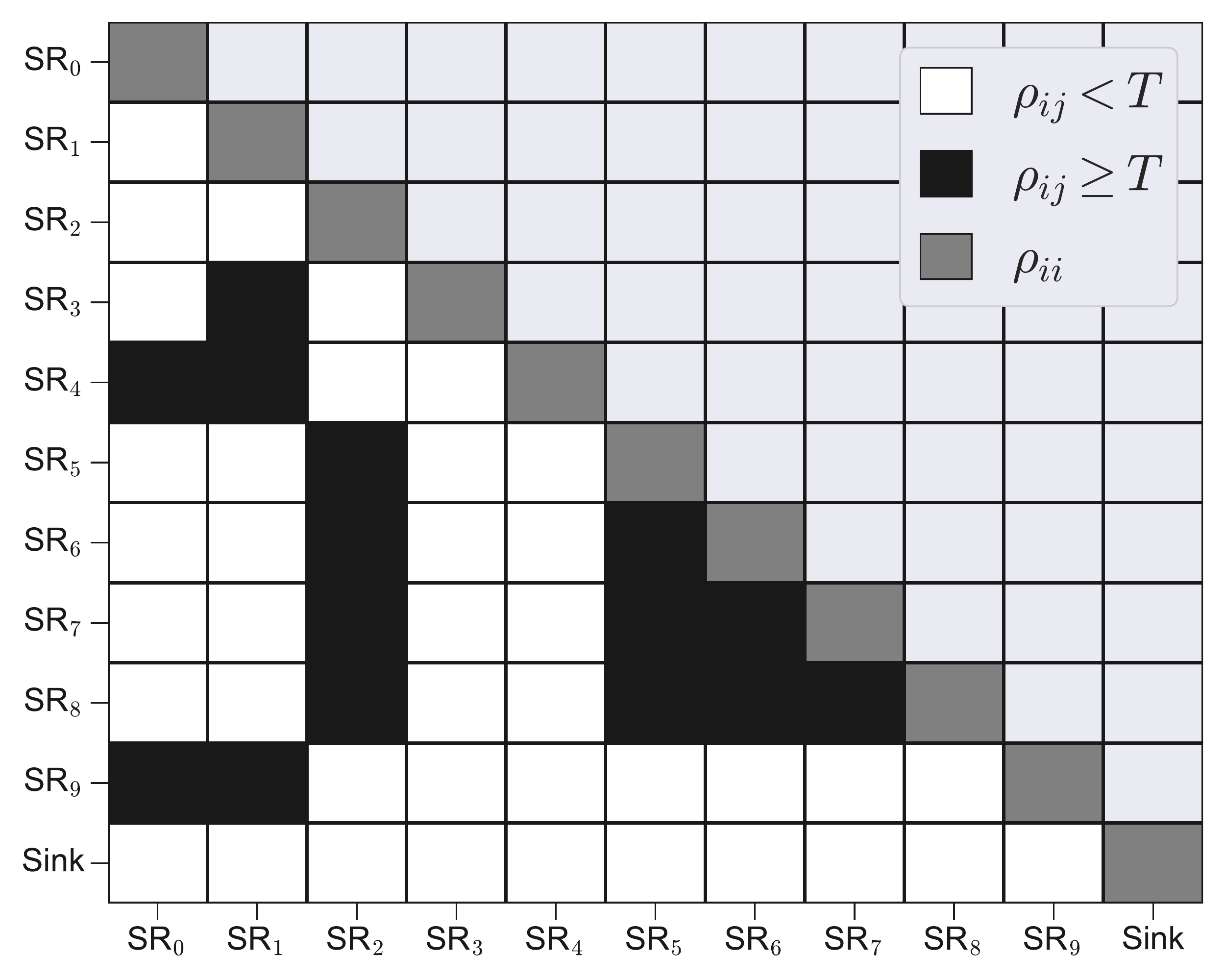}
  \caption{Overlap matrix of 10 signal regions ($\SR_0 - \SR_9$) with values
    masked according to threshold $T$ giving the exclusivity matrix. The final
    \emph{Sink} signal region has been inserted to provide a target for the
    path-finding algorithm}
  \label{fig:Path_finder 1}
\end{figure}

Figure~\ref{fig:Path_finder 1} shows the exclusivity matrix $E_{ij}$ of ten
signal regions, computed as the overlap matrix masked with a threshold $T$. The
matrix elements $\rho_{ij}$ that fall below $T$ are shown as white, and those
that are above are shaded black. For reasons that will become clear later, we
restrict ourselves to constructing combinations by adding SRs to the subsets in
strictly increasing index order.  Starting in the top left-hand corner of
Figure~\ref{fig:Path_finder 1} (at element $\rho_{00}$, or ($\SR_0,
\SR_0$)), the signal regions available for combination with $\SR_0$ are limited
to those corresponding to the white elements in the first column,
\ie~$E_{i,0} = 1$. We define $A_i$ as the ordered set of all \emph{allowed}
(non-overlapping) \SRs indices with respect to \SR$_i$ SR indices such that
 \begin{equation}
    A_i \equiv \{j: \rho_{i,j} < T, \; i < j < n\} \, .
    \label{eq:allowed_set0}
\end{equation}
Using Figure~\ref{fig:Path_finder 1} as an example, $A_0$ would be $\{1, 2, 3,
5, 6, 7, 8\}$. We can now expand on the previous definition of $K$, specifying
$K_{i}$ as a set of all allowed paths with initial elements $\SR_i$ such that:
\begin{equation}
  K_i \equiv \{ \{\SR_i, \ldots, \SR_\mathrm{final}\}, \ldots\} \, ,
  \label{eq:graph_set1}
\end{equation}
In this construction $K_{i, j}$ would be the $j${th} path within $K_i$, and by
extension $K_{i,j,k}$ would refer to the $k${th} element of $K_{i, j}$.
Applying this formulism to Figure~\ref{fig:Path_finder 1} and initiating a
subset $K_{0, 0, 0}$ with $\SR_0$, the available options for the second element
are given by indices in $A_0$ (equation~\eqref{eq:allowed_set0}). It follows
that $K_{0, 0, 1} = \SR_1$ as this is the first index in $A_0$, and thus $K_{0,
  0, 2} = \SR_2$ as this is the first available SR-index that is allowed by the
intersection of $A_0$ and $A_1$. Repeating the procedure and taking the
intersection of $A_1$ and $A_2$ results in an empty set, meaning that $K_{0,0}$
is a complete subset of three signal regions with overlaps below $T$. The next
combination, $K_{0,1}$, is the first allowed alternative to the final element of
$K_{0,0}$: $\{\SR_0, \SR_1, \SR_{2}\}$ becomes $\{\SR_0, \SR_1, \SR_{5}\}$.

This method of building paths is close to that of a depth-first search through
an unweighted directed acyclic graph where the ``nodes'' correspond to
signal-regions and ``edges'' to the allowed pairwise SR-combinations. The
directed and acyclic nature of the graph is enforced by the ordering of SRs and
the edges always pointing from lower to higher indices. However, there is a
major difference in that the choice of each signal region is dependent on those
allowed by all previous signal regions in the path, or in other words the
allowed vertices would be inherited. Fortunately this \emph{hereditary
condition} can be easily inserted into established DAG ``simple path''
algorithms~\cite{Sedgewick:2002, SciPyProceedings_11}.

Recasting the problem as an optimum-path search requires a few minor changes to
the definitions covered so far. Firstly, each path has to be defined between two
points: a source and a sink. As previously stated, each combination within the
subset $K_i$ has a defined source, however, the final signal region will depend
on the path taken. A convenient way of dealing with this condition is to define
a universally allowed $n$th signal region such that every possible path
terminates at index~$n$. This can be done by appending an $n$th ``sink'' signal
region to $\rho$, this is shown in Figure~\ref{fig:Path_finder 1} but can also
be expressed as
 \begin{equation}
   \rho_{n,i} = \rho_{i,n} = 0.0\; : \;  0 \leq i \leq n \, .
   \label{eq:M_set1}
\end{equation}
This modification of $\rho$ necessitates that the definition of $A_i$ also be
modified to include the $n${th} term:
\begin{equation}
  A_i \equiv \{j: \rho_{i,j} < T, \; i \leq j \leq n\} \, .
  \label{eq:allowed_set2}
\end{equation}

With $A_i$ defined in terms inclusive of $n$, we can define a modified
hereditary depth-first search (HDFS) algorithm that generates all the available
paths starting from an initial signal region.  This algorithm proceeds by
recursively appending diminishing subsets of allowed SRs $\mathcal{S}$, with
the current subset $\mathcal{S}_c$ defined as the the intersection of $A_c$
with the previous subset such that
\begin{equation}
    \mathcal{S}_{c} \equiv A_{c} \cap \mathcal{S}_{c-1} \, .
    \label{eq:completions}
\end{equation}
The remaining compatible SRs are hence given by the total intersection of the
compatible-SR sets for the elements already in the path. As this is constructed
iteratively, each stage of completion-refinement needs only to be compared
against the set of completions for the current final element,
$\mathcal{S}_{c-1}$). The HDFS algorithm uses this condition to efficiently
exclude overlapping SR-combinations from consideration.

\begin{figure}
  \centering
  \includegraphics[width=0.60\textwidth]{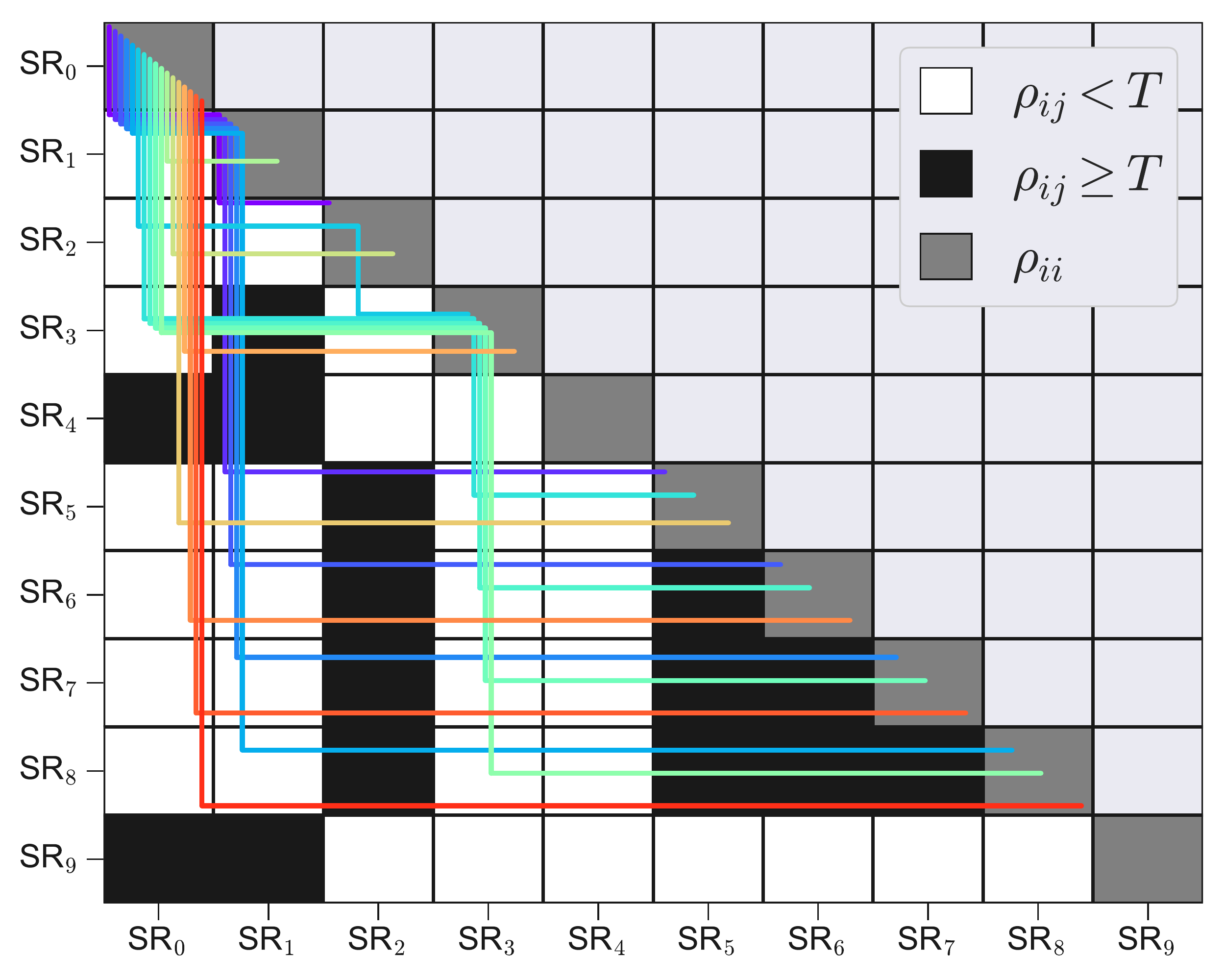}
  \caption{Exclusivity matrix of 10 signal regions ($\SR_0 - \SR_9$) with values
    masked according to threshold $T$. For clarity, the sink SR is not
    shown. The coloured lines show all allowed paths originating at $\SR_0$.}
  \label{fig:Path_finder 2}
\end{figure}

In summary, initiated from a source $\SR_i$, with the first element of
$\mathcal{S}$ being $A_i$, the set of all allowed paths $K_i$ can be built by
recursively evaluating the subsets of $\mathcal{S}$. Once the current iteration
has reached the ``sink'' $\SR_n$, a full path is defined by the steps
taken.
Figure~\ref{fig:Path_finder 2} shows the results from running this algorithm
using the exclusivity matrix from Figure~\ref{fig:Path_finder 1}, for paths
starting from $\SR_0$. The full DAG HDFS algorithm is given in pseudocode as
Algorithm~\ref{alg:DHAG} in Appendix~\ref{app:alg}.

\subsection{Weighted edges for sensitivity optimisation}
\label{sec:weightededges}

In generating the set of allowed paths, we have been concerned only with
SR-exclusivity and treated all graph edges (and hence SRs) as of equivalent
value, within the fixed DAG ordering provided. But in our physics application,
of course, this is not the case: for each specific BSM model, some signal
regions will be more sensitive than others. For example, leptophilic models
naturally tend to see most sensitivity in SRs with multilepton signatures;
models with enhanced couplings to the third generation have most impact on $t$-
and $b$-quark and $\tau$-lepton signatures; and dark-matter models favour jet +
missing transverse-energy signatures. In addition, when not all SRs have the
same integrated luminosity, SRs in high-luminosity datasets are naturally more
sensitive than those in low-statistics ones. These intuitive sensitivity metrics
can be incorporated into the graph model in the form of variable edge-weights.

Such weights should be motivated by the statistical goal being tested, and
ideally should be additive so standard longest-path optimisation can be used to
identify the most sensitive allowed SR-combination. A typically appropriate
choice for the edge weights, and the one used in this paper, is the logarithm of
the expected likelihood-ratios (LLR) between the signal-model under test and the
background-only model, $\ln(L_\mathrm{sb} / L_\mathrm{b})$, for pseudodata equal
to the expected yields under the background-only model. This is motivated by the
following logic:
\begin{enumerate}
\item As we are combining a set of direct-search analyses in which no individual
  significant signal was found, we choose to frame our mission primarily as
  maximising the volume of model-exclusion rather than a discovery. Our null
  hypothesis is hence the BSM signal model, and we seek to overturn it with a
  preference for the SM at every point in its parameter-space.
\item We hence aim to maximise the expected significance of exclusion
  $Z$ at each point in the BSM parameter space. Under the assumptions of Wald's
  Theorem~\cite{Cowan:2010js}, the expected significance is given by the
  square-root of the LLR between the models, hence maximising the LLR maximises
  the expected model-exclusion.
\end{enumerate}

As any generated path is by definition composed of signal regions which can be
treated as non-overlapping, the total log likelihood-ratio (LLR) $\sum_{i \in
  \mathrm{SRs}} \ln(L_{\mathrm{sb},i} / L_{\mathrm{b},i})$ of an SR subset is
just the sum of such weights along its corresponding path candidate. The use of
expected background pseudodata rather than the actual observed data-counts is
important to avoid cherry-picking of statistical fluctuations: we identify the
optimal SR-combinations for each point as if the data has not yet been recorded,
to avoid bias. Other use-cases, in particular anomaly-detection, in which the
observed data is compared to background expectations in search of the most
consistent, discrepant non-overlapping subset of measurements, require a
modified metric but with similar motivation. For such use cases, however, the
edge-weights are in general no longer additive, resulting in a more complicated
and CPU-intensive task.

In general, the optimal path can be found in reasonable time by evaluating the
overall sensitivity metric for every allowed SR-combination identified by the
HDFS algorithm of the previous section. However, in the case of additive
weights, further algorithmic optimisations are possible by a)~ordering the SRs
in decreasing order of individual sensitivity, and b)~exiting early from
generation of allowed-path subsets for which there is no possibility of
exceeding the metric obtained for the current maximum-sensitivity path. The
first of these conditions is simply implemented by \emph{a priori} ordering the
SRs according to decreasing expected LLR, such that paths containing the
expected dominant contributions to total LLR are evaluated first --- this opens
the possibility of evaluating only the sets of paths starting with the first
$\mathcal{O}(10)$ SRs. The second, however, makes such a manual cutoff largely
redundant by maintaining records of the highest complete-path LLR, and the sum
of LLRs over all remaining SRs in $\mathcal{S}_c$ as the allowed paths are
generated. Should the sum of the current path's LLR and its maximum possible
completion become smaller than the current best complete path, there is no point
in continuing to evaluate that set of completions and they can be
``short-circuited'' to further reduce the algorithmic complexity of the
path-finding.


We refer to this combination of DAG hereditary depth-first search and these
optimisations for weighted graph edges as the weighted HDFS algorithm (WHDFS).
This algorithm is our final method for efficiently addressing the specific
problem of finding the combination of statistically non-overlapping SRs which
maximises their additive combination of expected LLR sensitivities to a given
BSM model.


\subsection{Performance}
\label{sec:perf}

\begin{figure}
  \centering
  \includegraphics[width=0.60\textwidth]{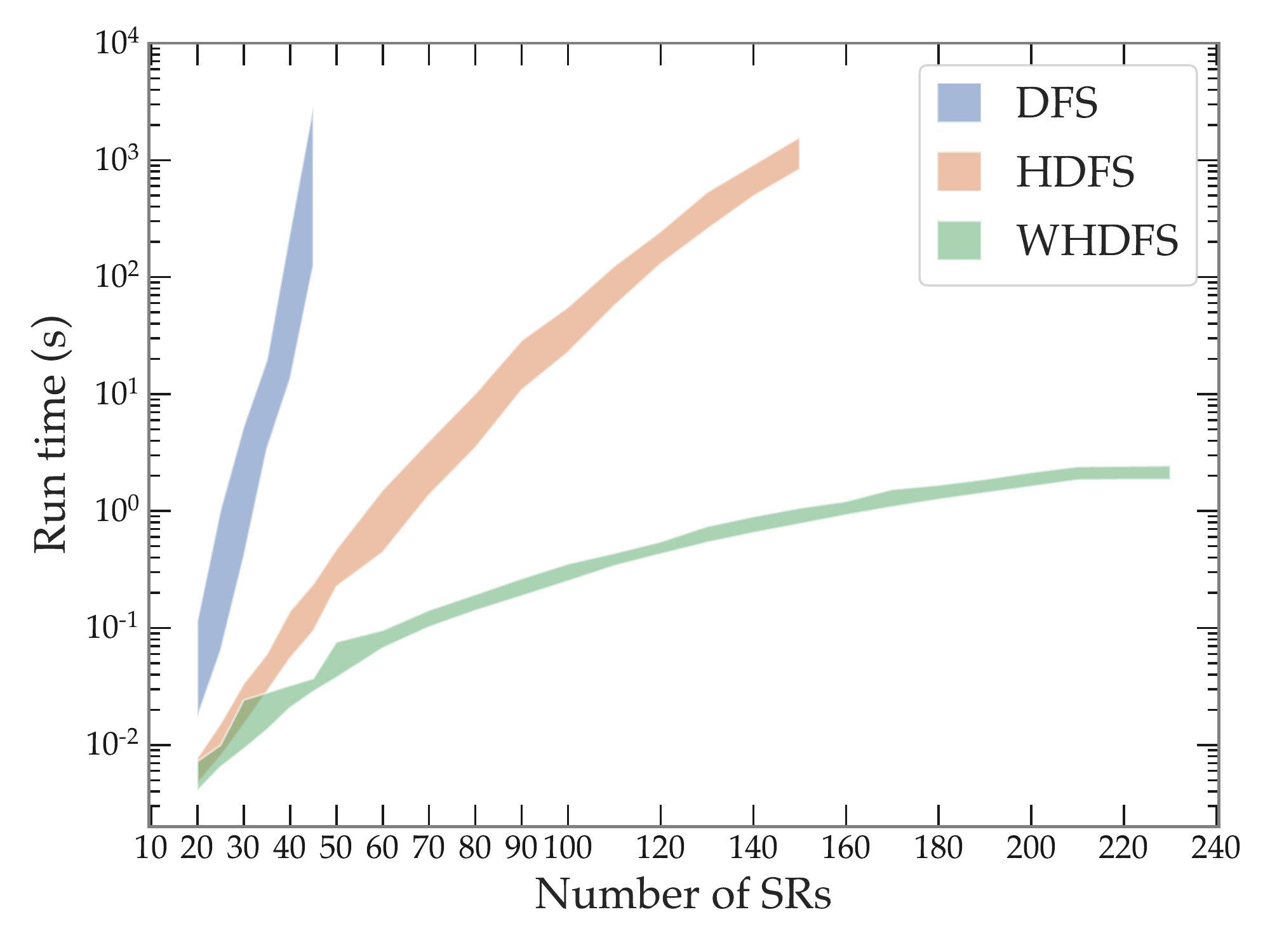}
  \caption{Comparison of CPU-runtime scaling against number of signal regions
    between the standard depth-first search (DFS), the hereditary DFS (HDFS),
    and the weighted hereditary DFS (WHDFS) algorithms.}
  \label{fig:comparison1}
\end{figure}

The algorithm performance was evaluated by randomly selecting 20 mass-points
from the ``T1'' simplified gluino-pair analysis to be shown in the following
physics-results section,
and calculating the optimal combinations for each of a set of reduced SR
collections $\{\SR_0, \ldots, \SR_{m-1}\}$ and its corresponding $m \times m$
exclusivity submatrix. The number of elements $m \le n$ in the reduced SR-sets
was evaluated from $m = 20$ to~80 in steps of~10, and from $m = 80$ to~140 in
steps of~20. The upper limit on $m$ was determined by the requirement to find 20
mass points (for timing-uncertainty estimation) with at least that many
supported SRs.

Figure \ref{fig:comparison1} shows the CPU-performance comparison between the
three graph-based algorithms discussed in Section~\ref{sec:pathfinding} on these
SR-combination problems of varying size. The plot clearly shows that the simple
depth-first search (DFS) does not scale sufficiently for physics purposes: in
BSM scans considering many thousands or millions of model points and hundreds of
SRs, the decision of which SR combination to use for each point needs to be made
typically on the order of seconds, but the DFS algorithm requires hundreds of
seconds by 40~SRs, with extremely strong exponential scaling. The HDFS algorithm
fares much better, scaling up to 100~SRs with a flatter exponential growth than
DFS, and with slightly sub-exponential thereafter. Regardless, it requires
$\mathcal{O}(100)$ seconds for 100~SRs, insufficient for many practical
applications. But the further optimisations enabled by the WHDFS formulation
show a flatter still scaling exponent, with sub-exponential growth that becomes
particularly flat for large SR counts. 230~SRs were obtained in around
2~seconds, very compatible with adaptive sampling, and indicating little issue
in scaling further toward thousands of SRs.
These performance gains indicate the effectiveness of the WHDFS algorithm and
that it can meet the current practical requirements of large-scale analysis
combination.


\section{Results}\label{sec:results}

To illustrate the power of our approach, we now present physics results for
various BSM-reinterpretation scenarios. In order of increasing complexity, we
first demonstrate increases in model-exclusion limits in the context of
simplified models in Section~\ref{sec:res:t1tttt}. Raising the stakes, we then
demonstrate the effect of our combinations on the pMSSM-19 model in
Section~\ref{sec:res:pmssm}.  We end with a discussion of our combination
results in the context of a simple $t$-channel dark matter model, this time
fully recasting the relevant analyses, in Section~\ref{sec:res:tcdm}.

As is the case throughout this paper, control regions are ignored, as are
overlaps in the background expectations of the signal regions. This reflects an
implicit assumption that the signal regions are specific enough to event
topology and kinematic phase-space that events falling into them are
indistinguishable between signal and background (the job of removing reducible
backgrounds having already been performed by the pre-selection and SR-cut
definitions). One could hence also perform the overlap estimation using large
background-event samples in place of the sampling over signal models.

\subsection{T1 simplified-model combination}
\label{sec:res:t1tttt}

Following the method described in Section~\ref{sec:correst}, an overlap matrix
was constructed from the selection of analyses available in \smodels and
\MA. Using the \smodels database, an exclusivity matrix of 393~\SRs was created
with a 1\% threshold of maximum overlap. The first test of the \taco formalism
was to compare the combination results to the validation plots for analyses and
topologies available in the \smodels database, chosen as this checks for
consistency within a model-space completely understood and mapped by
\smodels. The first simplified model chosen was the ``T1'' topology, which is a
simplified version of gluino pair production in which each gluino undergoes a
three-body decay $\Tilde{g} \rightarrow q\bar{q}\Tilde{\chi}_1^0$ to a
light-flavor quark-antiquark pair plus the lightest stable particle (LSP)
$\Tilde{\chi}_1^0$.

\begin{figure}[t]
    \centering
    \subfloat[Expected]{{\includegraphics[width=0.48\textwidth]{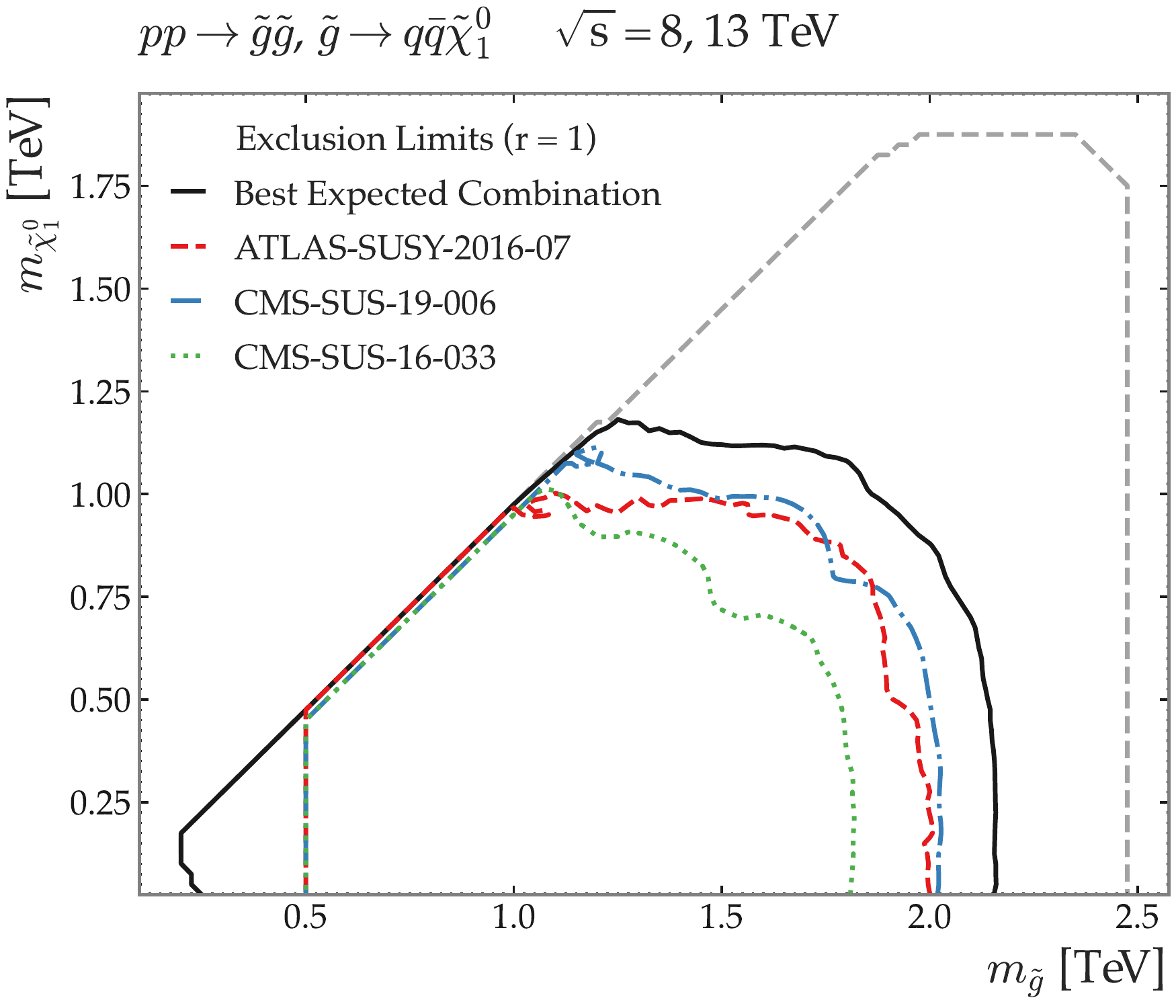} }}
    \subfloat[Observed]{{\includegraphics[width=0.48\textwidth]{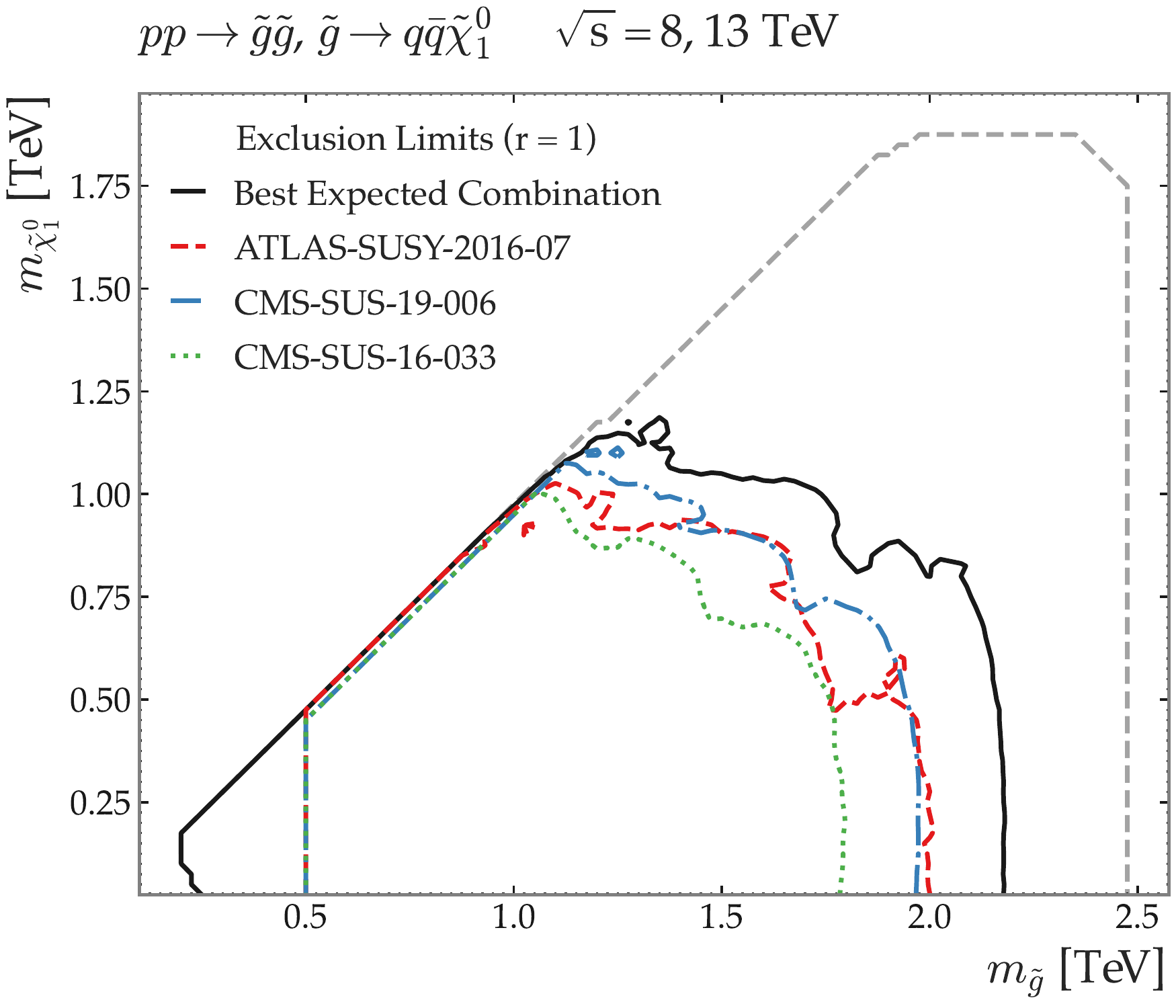} }}
    \caption{Validation plots comparing the (a)~expected and (b)~observed
      results of the \taco SR-combination against three individual-analysis
      limits available in the \smodels results database for the T1 topology. The
      lines show the exclusion limits in terms of $r$-value where $r = 1$ is
      analogous to the 95\% confidence-level exclusion. The CMS-SUS-19-006-ma5
      analysis is so-named due to the efficiency map being obtained for \smodels
      by use of \MA rather than direct from the experiment-provided analysis
      data. The dashed-grey line indicates the boundary of the efficiency maps.}
    \label{fig: T1 Results}
\end{figure}



Figure~\ref{fig: T1 Results} shows the T1-topology validation plot of
(a)~expected and (b)~observed limits for the combined SRs against three
individual analyses. The contour lines show the exclusion limits in terms of the
ratio of the predicted cross-section $\sigma_{\mathrm{pred}}$ and the upper
limit on that cross-section $\sigma_{\mathrm{UL}}$ , $r \equiv
\sigma_{\mathrm{pred}}/\sigma_{\mathrm{UL}}$, such that $r = 1$ corresponds to
the line of exclusion at 95\% confidence. A total of 265 SRs were available with
contributions to the T1 topology. For each point in the model space, the number
of available SRs was determined by identifying those with efficiency maps
whose parameter ranges included the model point. Once a set of available SRs was identified, they
were ordered by the expected upper-limit (UL) on the expected yield (luminosity
$\times$ cross-section $\times$ efficiency) at the model point for each SR. This
selection and ordering of the signal regions was propagated to the exclusivity
matrix, and the WHDFS SR-selection algorithm was applied. Figure~\ref{fig: T1
  Results} shows that the combined result pushes the exclusion line beyond that
of the best performing analysis available in the current \smodels database by
approximately \SI{150}{\GeV}. 

\begin{figure}[t]
    \centering
    \subfloat[Initial node in each combination]{{\includegraphics[width=0.49\textwidth]{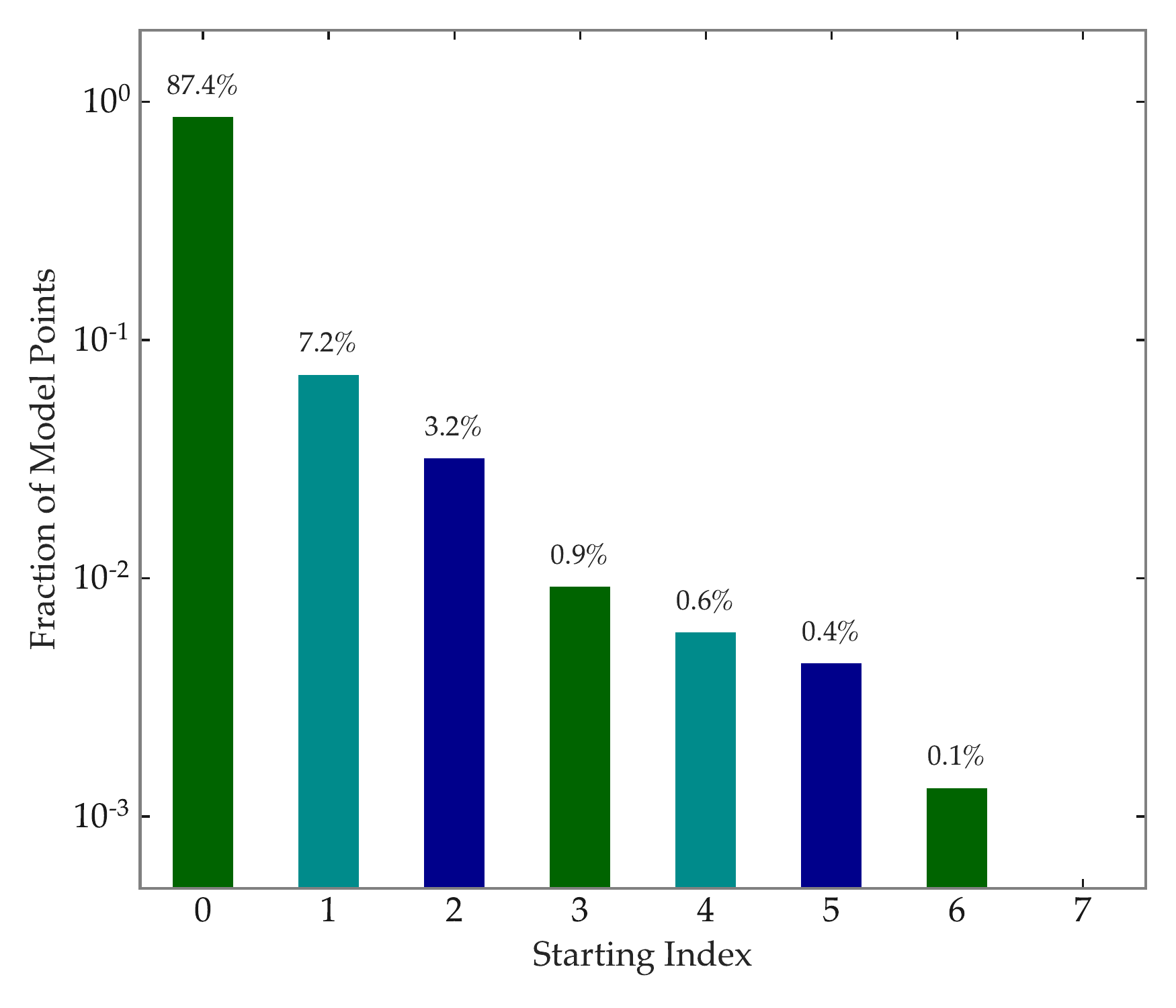}}}
    \subfloat[Number of \SRs in combination]{{\includegraphics[width=0.49\textwidth]{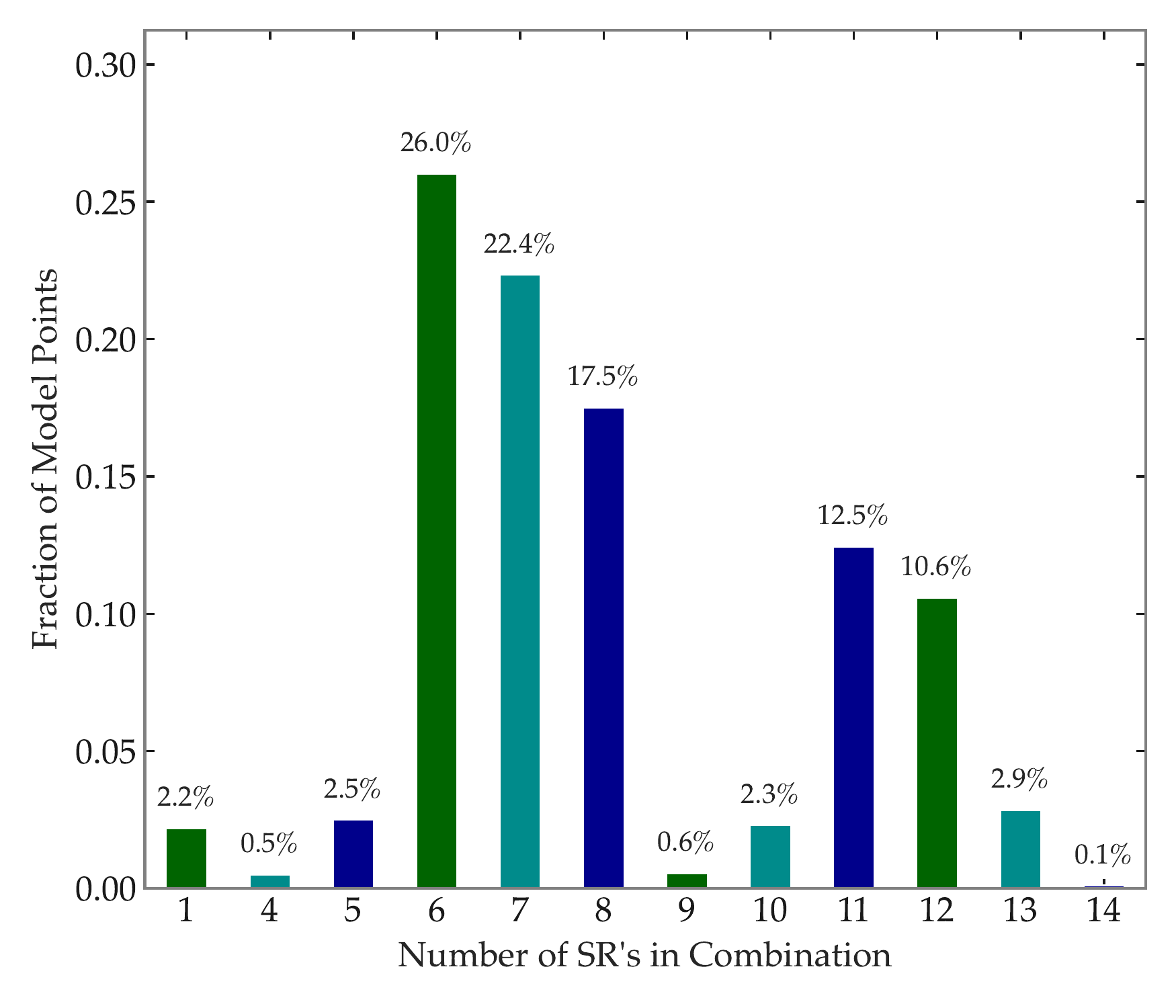} }}
    \caption{Fractional distributions of (a)~starting SR-index and (b)~number of
      SRs in each combination. Data is taken from the optimum SR
      combination found for each model point in the T1 model space. As mentioned
      in Section~\ref{sec:srcomb} the nodes are constructed from an ordered set
      of SRs that is optimised for each point in the model space.}
    \label{fig: T1 review}
\end{figure}

Looking deeper into the combined results, Figure~\ref{fig: T1 review}(a) shows
the distribution of starting (\ie~lowest) SR-indices over the set of
maximum-sensitivity combinations. This confirms the statements made in
Section~\ref{sec:ranking} that the efficiency of the path-finding would be greatly
increased by sorting the exclusivity matrix by individual SR sensitivities. The
histogram shows that, when ordered, the optimum combination is typically seen
early in the iteration process, allowing many later path-sets to be vetoed when
there is no prospect of their completions beating the current best. The
right-hand plot of Figure~\ref{fig: T1 review} shows the percentage prevalence
of the number of SRs in each best-sensitivity combination, with typically 6--10
of the available 265~SRs being used. This small number is conveniently
compatible with expensive statistical treatments such as coherent profiling or
marginalisation of systematic uncertainties across analyses, which would be
prohibitively expensive over the 265 (and ever-increasing) full set of SRs.


\begin{figure}[t]
    \centering
    \subfloat[Expected]{{\includegraphics[width=0.475\textwidth]{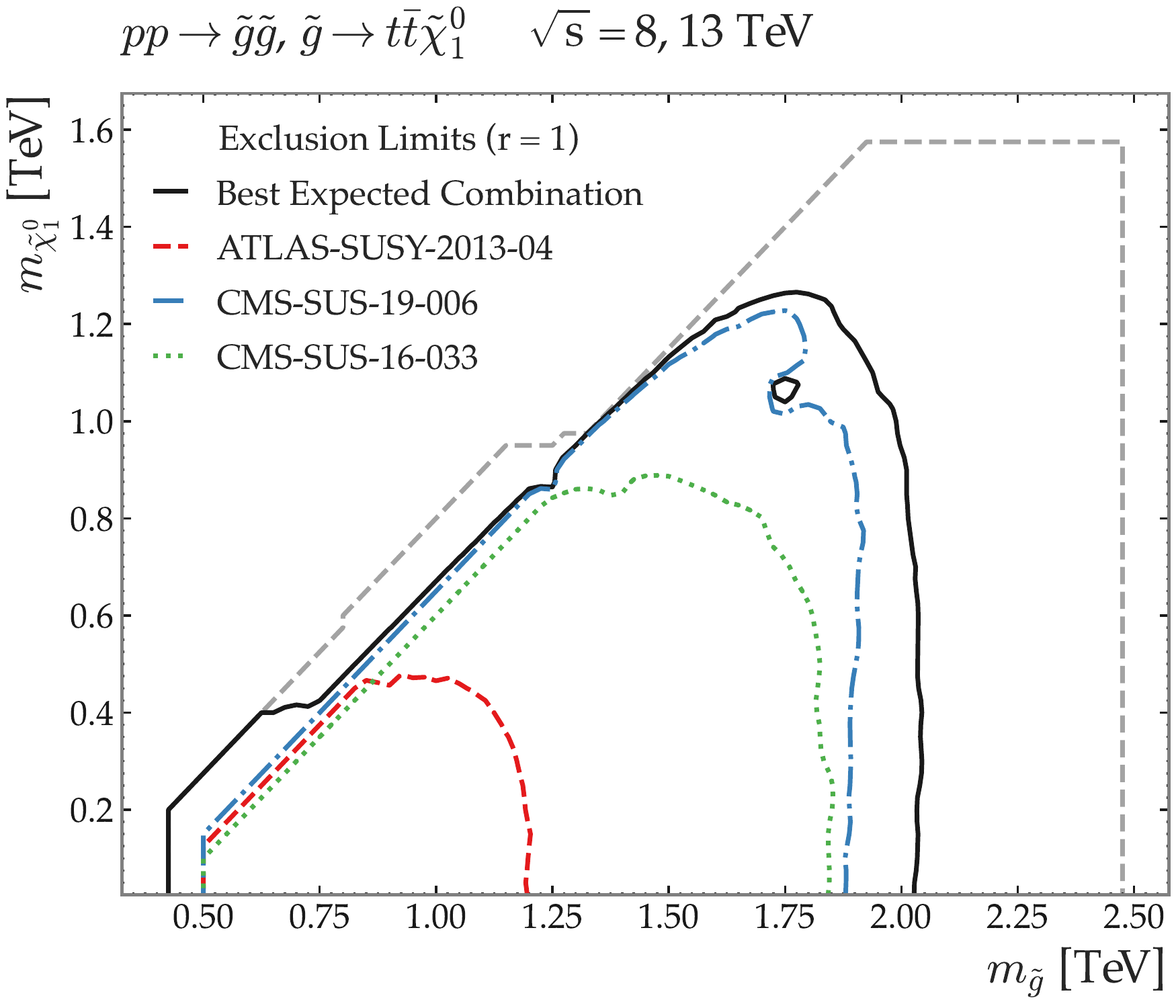} }}
    \subfloat[Observed]{{\includegraphics[width=0.475\textwidth]{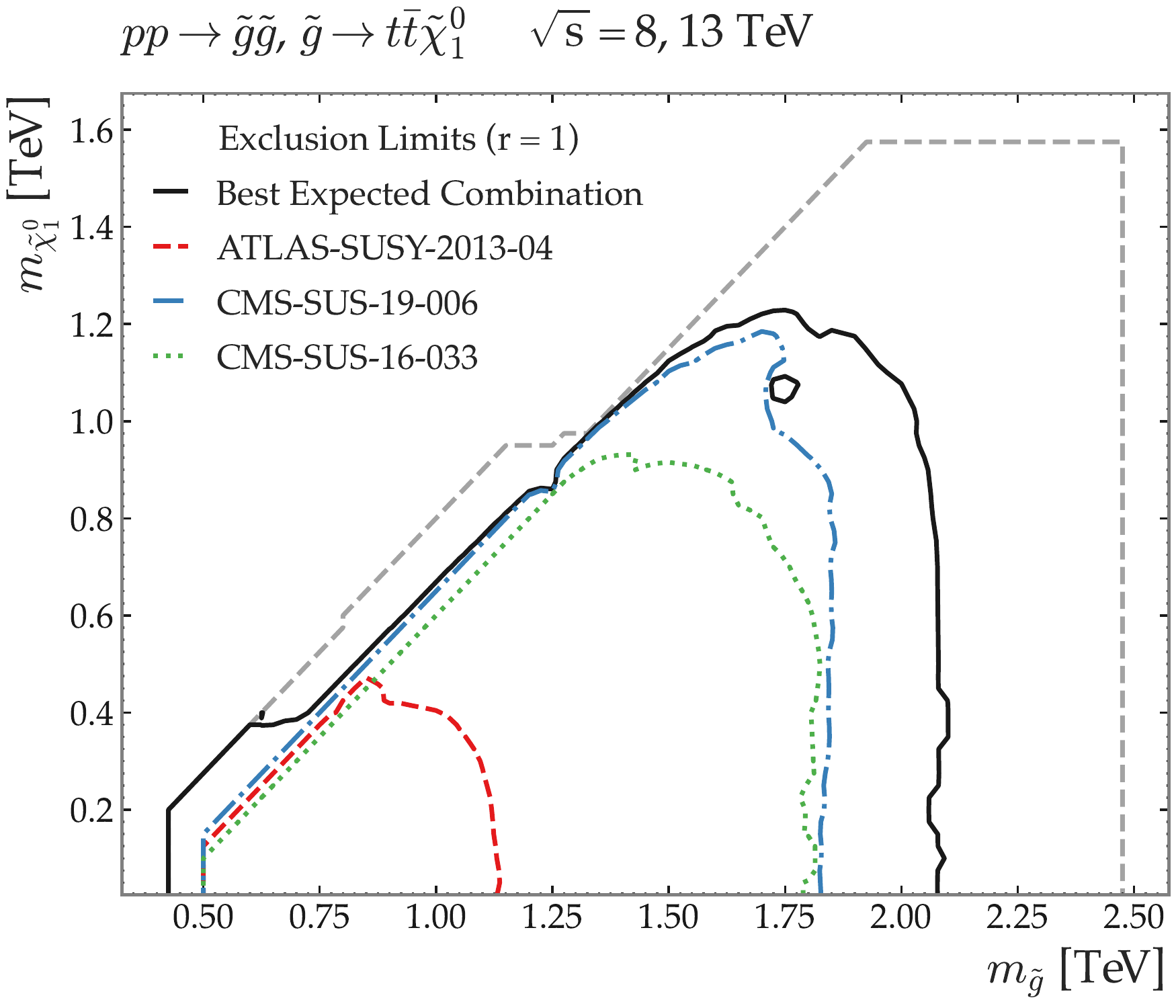} }}
    \caption{Validation plots comparing (a)~the expected and~(b) observed
      results of the combination results with 3~single analyses available in the
      \smodels results database for the T1tttt topology. The lines show the
      exclusion limits in terms of $r$-value where $r = 1$ is analogous to the
      95\% confidence level. The dashed-grey line indicates the boundary of the
      efficiency maps.}
    \label{fig: T1tttt Results}
\end{figure}

We also considered the T1tttt topology, a modification of the T1 model in which
the gluino decays exclusively into top quark--antiquark pairs ($\Tilde{g}
\rightarrow t\bar{t}\Tilde{\chi}_1^0$). Figure~\ref{fig: T1tttt Results} shows
(a)~the T1tttt topology validation plot of expected and (b)~observed results,
again comparing the exclusion ranges of the combined SRs to that of three
individual analyses. The construction of the plot follows the same methods used
for Figure~\ref{fig: T1 Results}. Similarly, the dominant contribution to the
T1tttt combinations is the CMS-SUS-19-006 analysis~\cite{CMS:2019zmd} (seen in
the blue dot-dashed exclusion line). Again, a significant expansion of the 95\%
exclusion contour over the single-SR results is seen, with the combination seen
to smooth out the particular weakness of the most constraining analysis around
$m_{\tilde{\chi}^0_1} \sim \SI{1.1}{\TeV}$.



\subsection{pMSSM-19 reinterpretation}
\label{sec:res:pmssm}

With the machinery in place to construct the exclusivity matrix from \SRs based
on a given model-point, it was now possible to extend the analysis to
increasingly complex models. The 19-parameter phenomenological Minimal
Supersymmetric Standard Model (pMSSM-19) was chosen as a testing ground for
reinterpretation, as a model with considerably more degrees of freedom than
typically studied in experimental publications. We used data points sourced from
the ATLAS 2015 pMSSM-19 scan paper~\cite{pMSSM}, independently for that paper's
two scenarios of bino-like and wino-like LSPs. These points were identified by
ATLAS according to their model viability compared to \SI{8}{\TeV} ATLAS data,
and hence provide an \emph{a priori} interesting set for re-evaluation against
\SI{13}{\TeV} LHC data.

$p$-values from \SI{13}{\TeV} LHC Run-2 data analyses
were calculated from
the first (randomly ordered) \num{27000} model points in the bino and wino
scenarios separately, run through the \smodels analysis chain and discarding
those points which lay outside the bounds of the \smodels efficiency maps, with
$\sim \! \num{20000}$ points surviving in each run.

\begin{figure}
    \centering
    \subfloat[Expected]{{\includegraphics[width=0.49\textwidth]{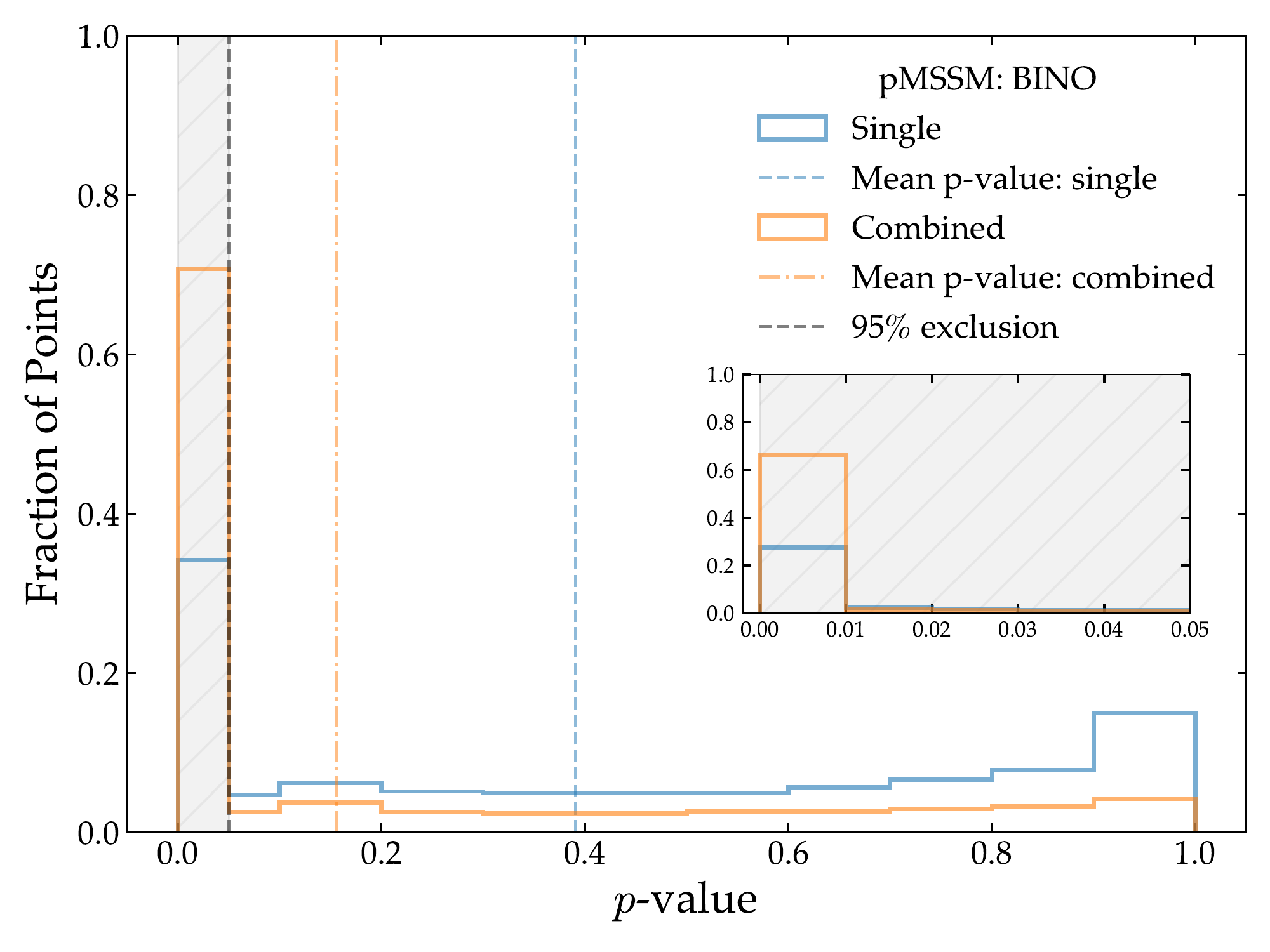} }}
    \hfill
    \subfloat[Observed]{{\includegraphics[width=0.49\textwidth]{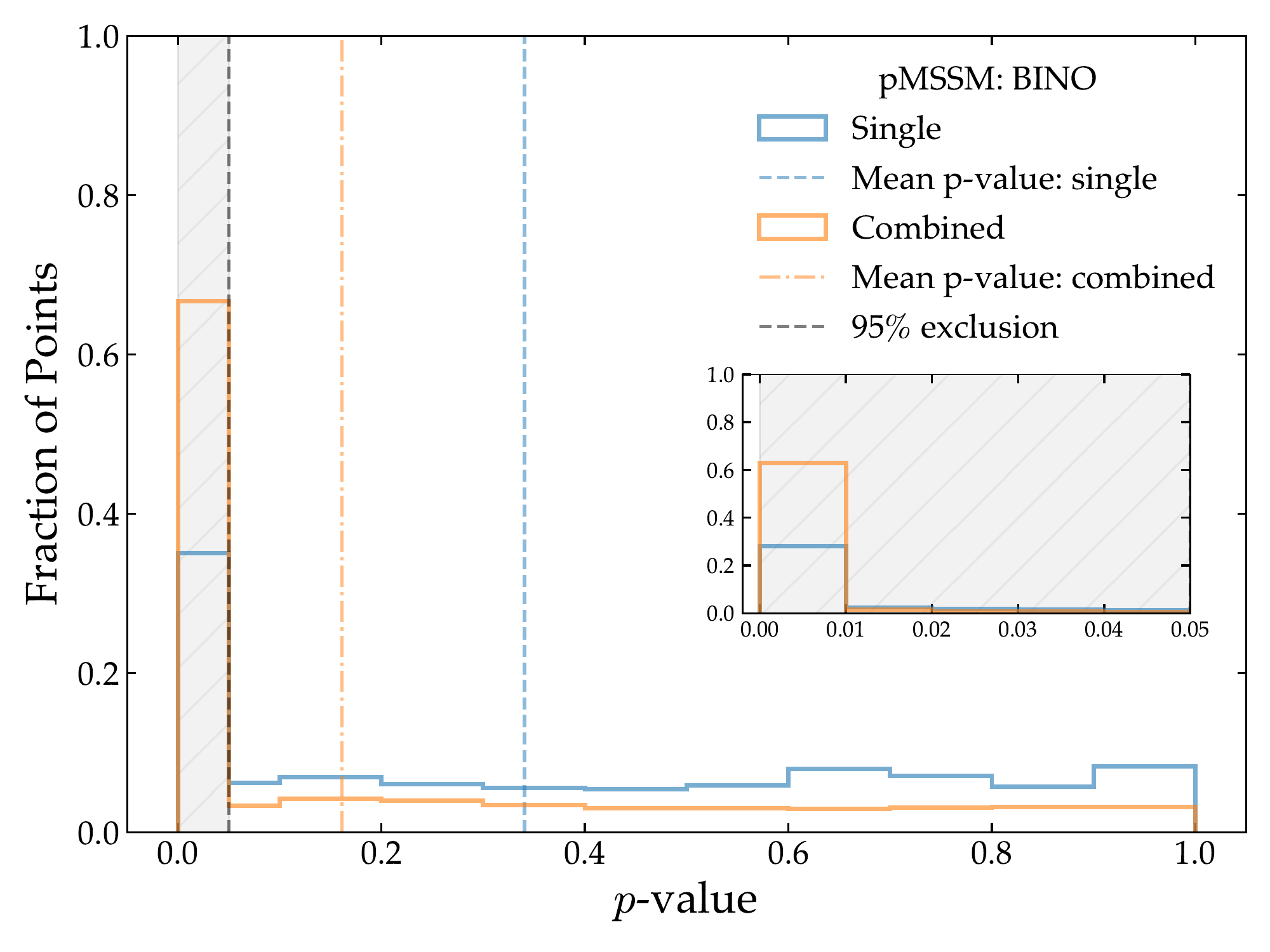} }}

    \caption{Results from the pMSSM-19 bino reinterpretation using the \taco
      combination method. $p$-values were calculated from a selection of
      \num{22000} points taken from the ATLAS pMSSM-19 data set.
      The blue and orange dashed lines show the mean $p$-values for
      the single and combined results respectively.
      The histograms show that in both the (a)~expected and (b)~observed
      cases a large fraction of points are moved beyond the 95\% exclusion limit
      by WHDFS SR-combination.}
    \label{fig: Bino hist}
\end{figure}

LLRs were calculated using the \smodels best-single-expected SR-selection
process and the best-expected-combination results. The resulting $p$-value
distributions from the pMSSM-19 bino-LSP reinterpretation are shown in
Figure~\ref{fig: Bino hist}. The histograms show that in both the (a)~expected
and (b)~observed cases the combination procedure moves a significant fraction of
points from below the 95\% exclusion into the excluded category, an increase in
exclusion fraction from approximately 35\% to 70\% of all points in the ATLAS
set of bino-like models.

\begin{figure}
    \centering
    \subfloat[Expected $P(\mathrm{combined} | \mathrm{single})$]{{\includegraphics[width=0.495\textwidth]{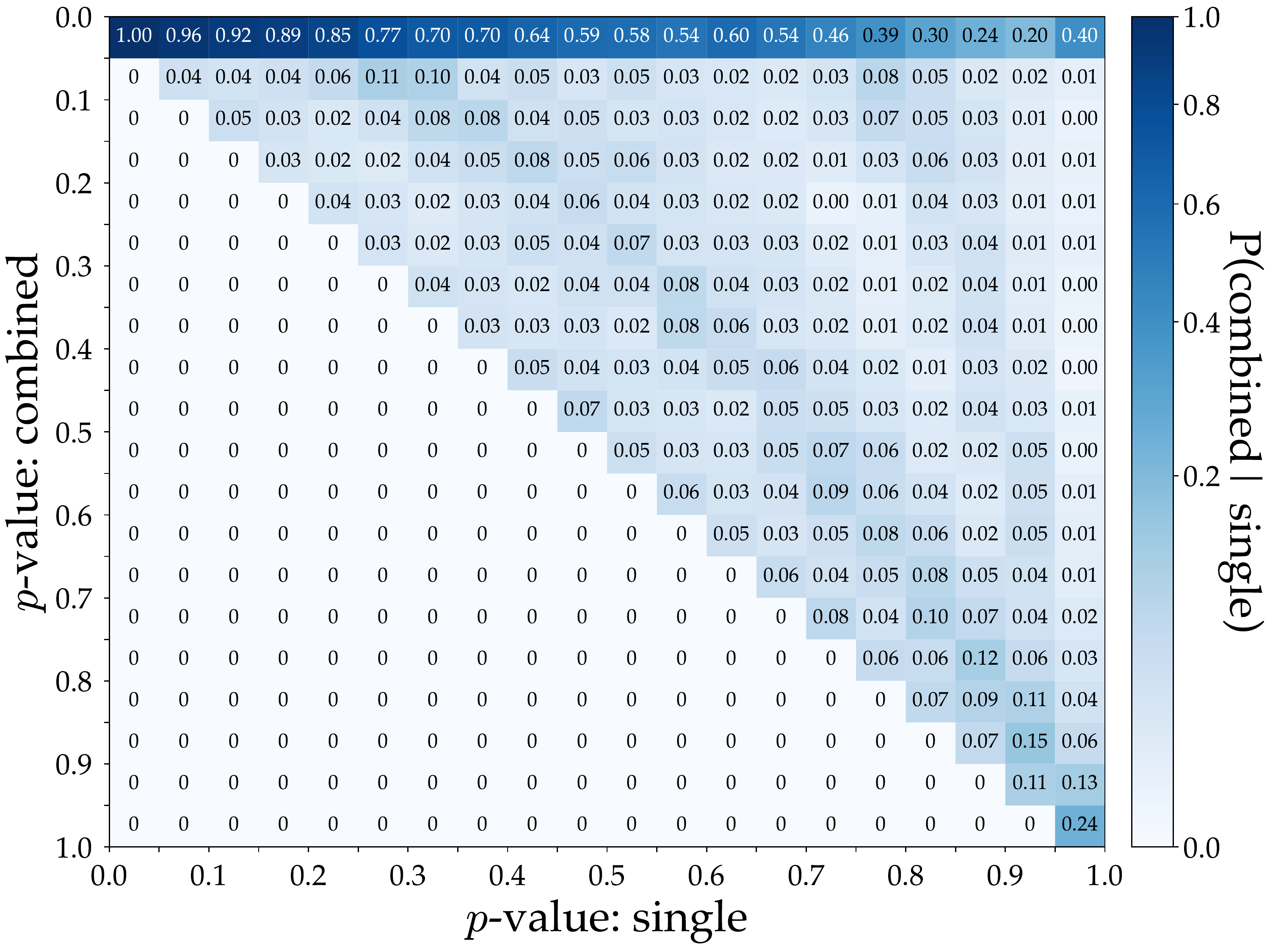} }}
    \subfloat[Expected $P(\mathrm{single} | \mathrm{combined})$]{{\includegraphics[width=0.495\textwidth]{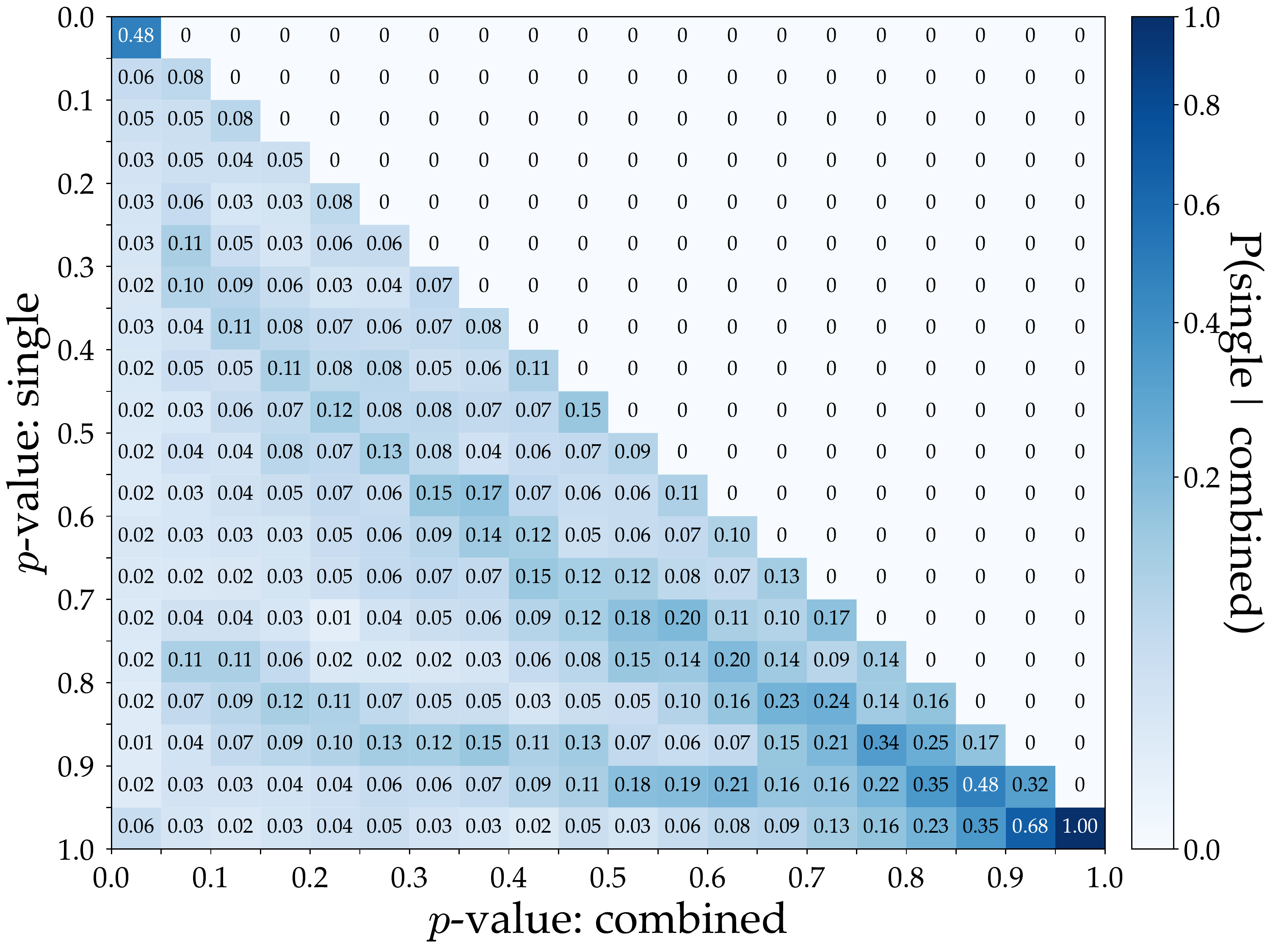} }}\\
    \subfloat[Observed $P(\mathrm{combined} | \mathrm{single})$]{{\includegraphics[width=0.495\textwidth]{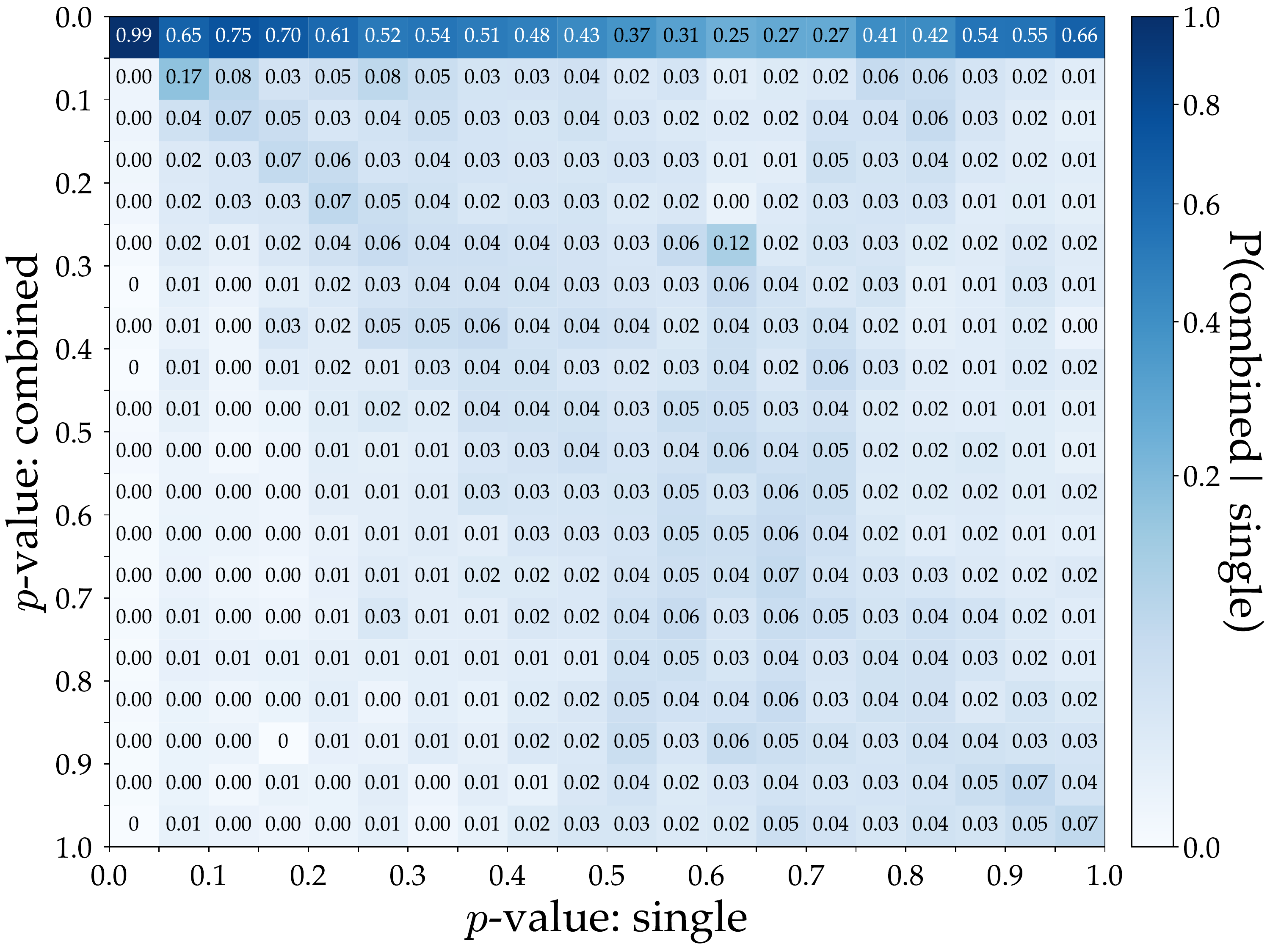} }}
    \subfloat[Observed $P(\mathrm{single} | \mathrm{combined})$]{{\includegraphics[width=0.495\textwidth]{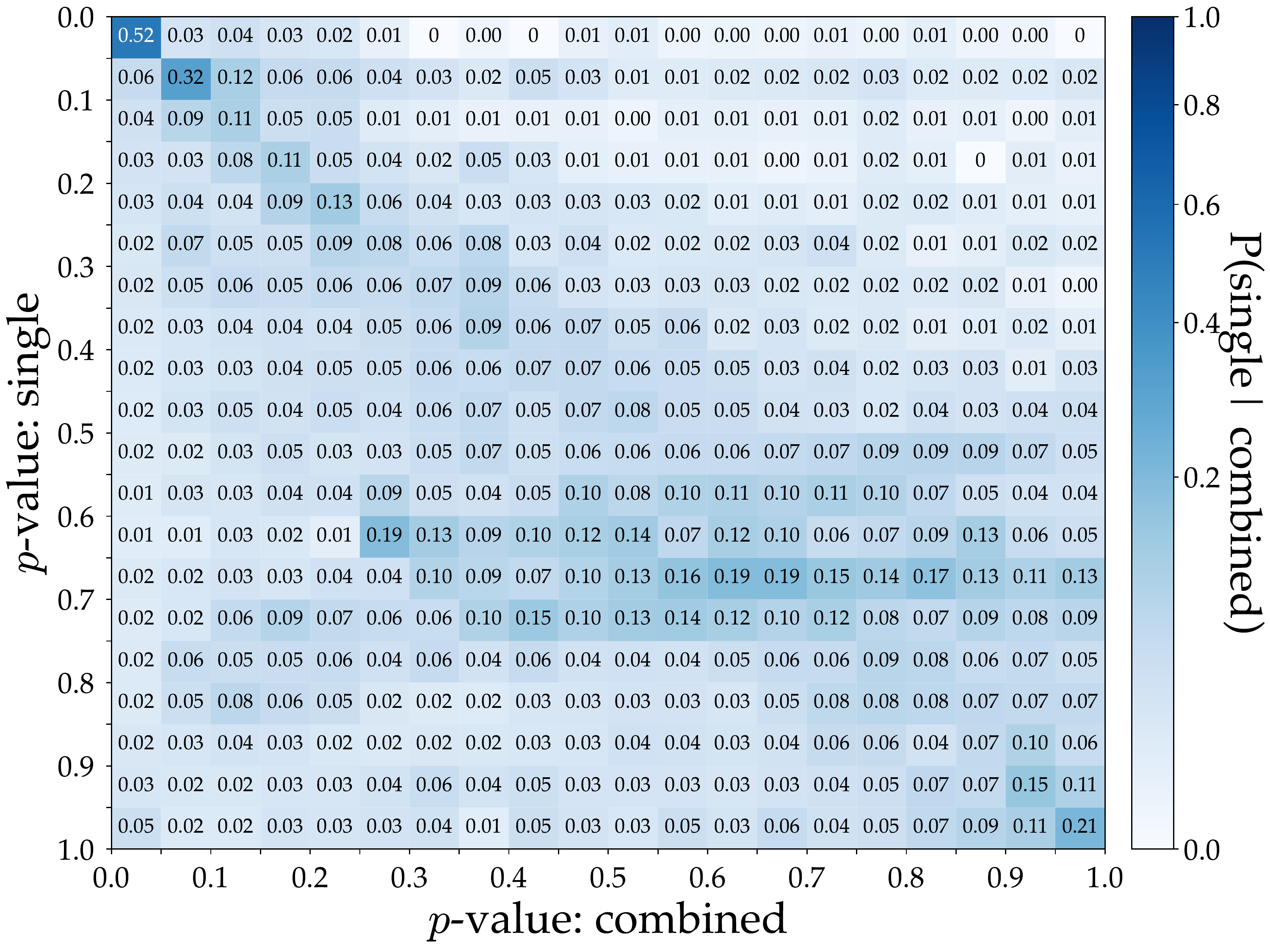} }}

    \caption{Transition matrices showing the pMSSM-19 bino results. The matrices
      describe the probability of a model point moving from one $p$-value bin to
      another, by use of the SR-combination scheme rather than the conservative
      single-best-SR strategy adopted by current recasting tools. The subfigure
      columns split the transition behaviours by the transition to combined
      $p$-value distributions given single-SR performance, and the single-SR
      origins of each combined-SR $p$-value range. The top and bottom subfigure
      rows show the expected and observed transitions respectively.
    }
    \label{fig: BINO Results}
\end{figure}

This mean shift from single to combined could be seen as confirmation of the
exclusionary power of the \taco approach, but before making any conclusions it
is prudent to review exactly how the model points are behaving on a bin-to-bin
level.  To this end, Figure~\ref{fig: BINO Results} shows the transition
matrices (also known as stochastic matrices) for the pMSSM-19 bino dataset,
showing the probability of a model point ``moving'' between $p$-value bins
depending on whether the single-SR or combined-SR LLR construction method is
used.  This can either be framed as the probability of points in a particular
single-SR bin moving to each combined-SR bin, or the ``origin distribution'' of
the points ending up in a particular combined-SR $p$-value bin; both versions
are informative and are shown in the left- and right-hand subfigure columns
respectively, with expected and observed results in the subfigure rows. As the
values in each matrix are given as a probability
$P(\mathrm{row}|\mathrm{column})$ the sum over the column values equals~1.

We start with the expected $P(\mathrm{combined} | \mathrm{single})$ result shown
in the top-left of Figure~\ref{fig: BINO Results}\,(a), which shows how \taco
combination changes the $p$-values of model points given their initial $p$-value
as obtained from the single best-expected SR. The overall form of the transition
pattern is logically consistent with the aim to use \SR combinations to
\emph{increase} exclusionary ability, thus it is expected that all transitions
be located in the upper triangle of the $P(\mathrm{combined} | \mathrm{single})$
matrix. Encouragingly, the transitions are dominated by movements into the
excluded $p \in [0, 0.05)$ bin, not just from neighbouring ``nearly excluded''
  single-SR bins, but across the whole spectrum of single-SR $p$-values: this
  shows that even 40\% of the least excluded single-SR points can be excluded
  when combination of independent SRs is enabled. Some subleading transition
  structures can be seen within the matrix, showing below-threshold increases in
  exclusion which can potentially be brought above threshold by availability of
  more analyses.  The expected $P(\mathrm{single} | \mathrm{combined})$ results
  in Figure~\ref{fig: BINO Results}\,(b) are concentrated in the lower triangle
  of the matrix, as expected, with similar evidence of structures in the
  transition pattern.

Moving to the observed case in the lower plots of Figure~\ref{fig: BINO
  Results}, the results become more nuanced as the distribution of transition
becomes dilute. The dominant transition into the exclusion bin identified in
plot~(a) is replicated in the observed case shown in plot~(c), as was expected
from the histogram results. The ``negative transition'' of model points in the
direction opposite to expectation is caused by over-fluctuations in the observed
yields of the \SRs used to calculate the combined result. This may be a
statistical feature intrinsic to combining multiple \SRs, although only present
in a small minority of cases ($\approx 5\%$). Looking back to Figure \ref{fig:
  Bino hist}\,(b) it can be seen that the percentage of points in the exclusion
bin jumps from $\approx$ 35\% to over 90\% when using the combined \SRs. Thus,
the negative transitions seen in plot~(d) of Figure~\ref{fig: BINO Results} are
only representative of a small fraction of model points.

\begin{figure}
    \centering
    \subfloat[Expected]{{\includegraphics[width=0.49\textwidth]{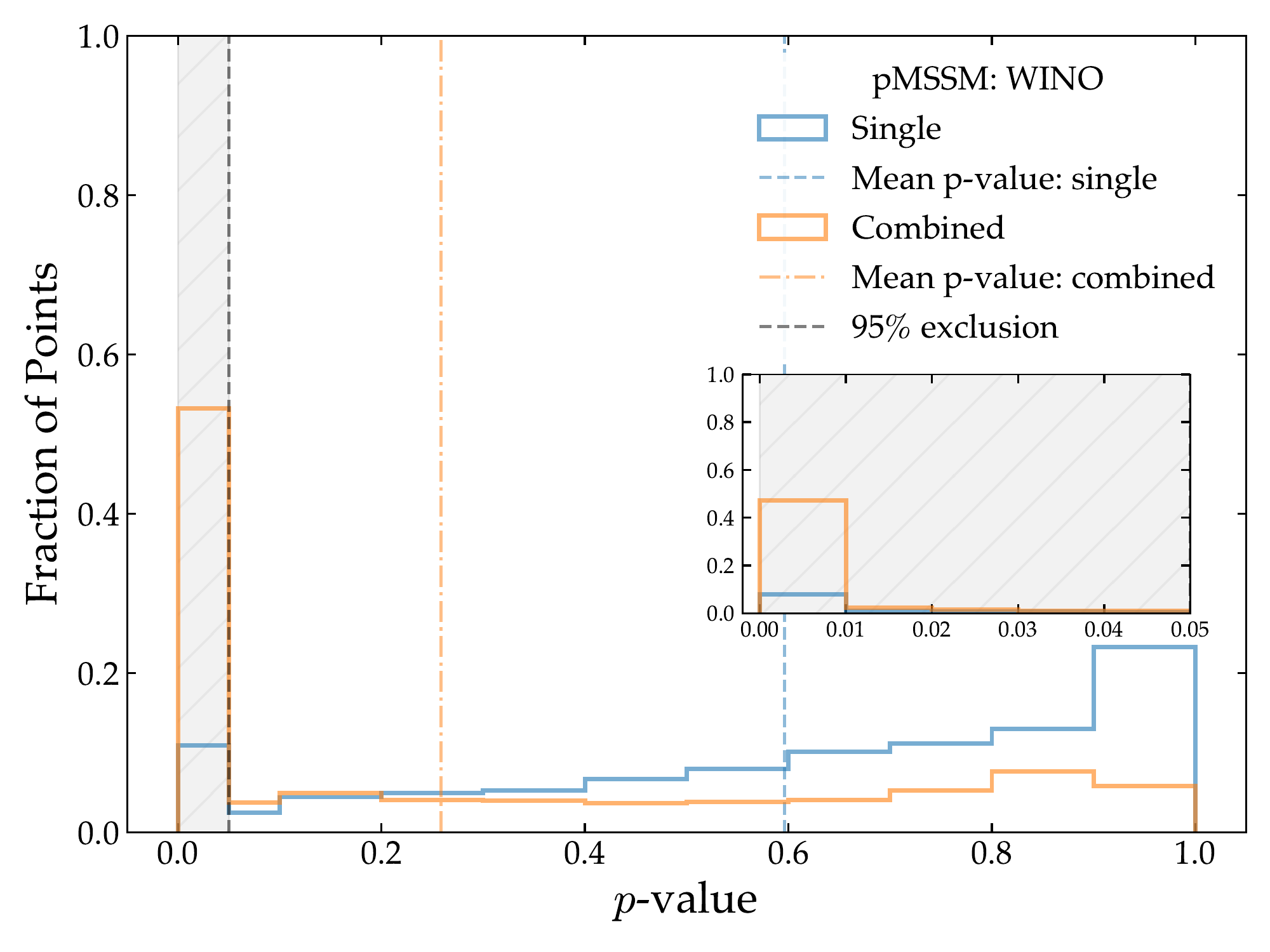} }}
    \hfill
    \subfloat[Observed]{{\includegraphics[width=0.49\textwidth]{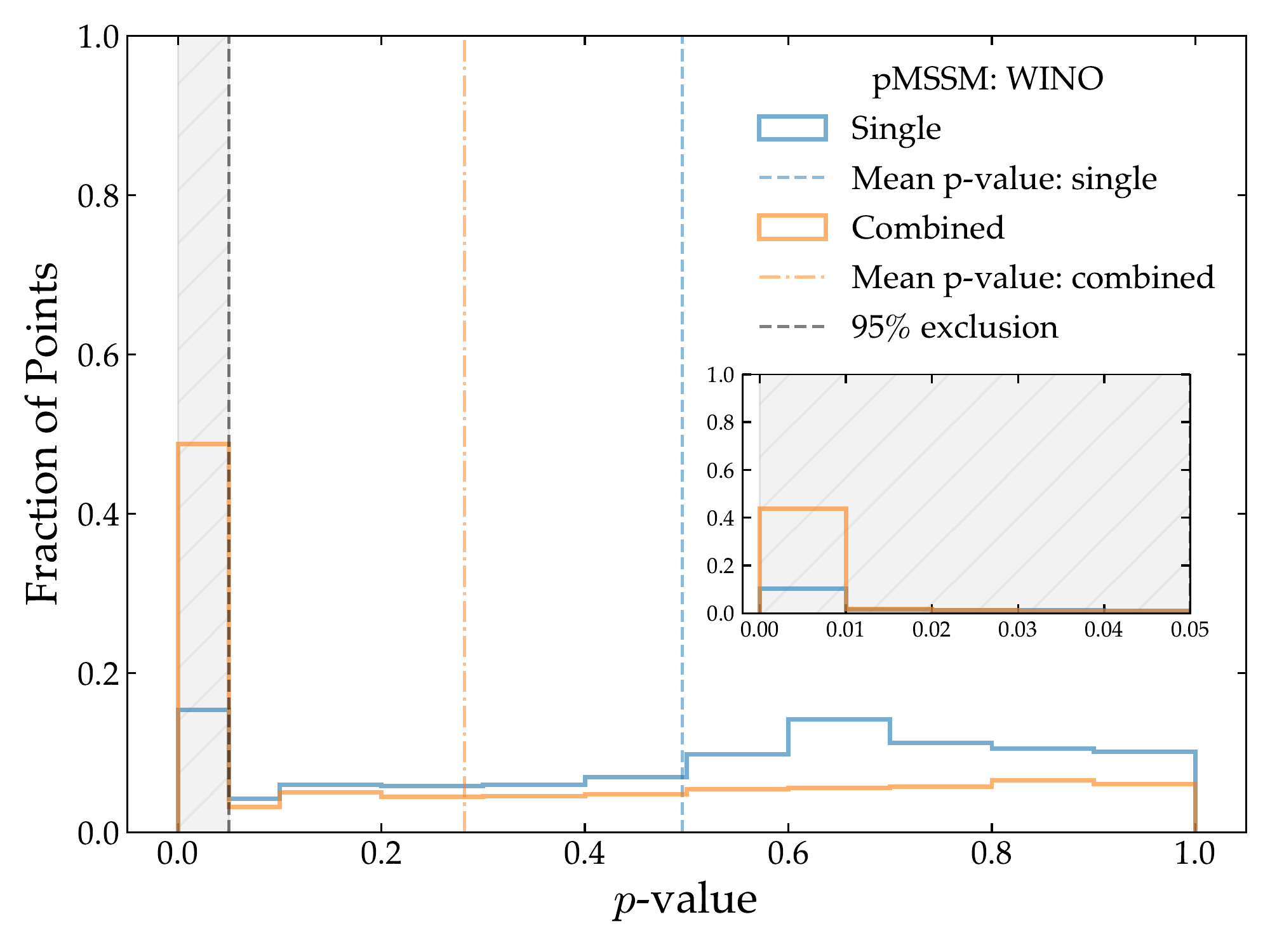} }}

    \caption{Results from the pMSSM-19 wino reinterpretation using the \taco
      combination method. $p$-values were calculated from a selection of
      \num{20000} points taken from the pMSSM-19 data set.
      The blue and orange dashed lines show the mean $p$-values for
      the single and combined results respectively.
      The histograms show that in both the (a)~expected and (b)~observed
      cases a large fraction of points are moved beyond the 95\% exclusion limit
      by WHDFS SR-combination.}
    \label{fig: Wino hist}
\end{figure}

Figure~\ref{fig: Wino hist} shows the pMSSM-19 wino-LSP reinterpretation
results. When compared to the bino results in Figure~\ref{fig: Bino hist} we can
see a similar overall shift to the exclusion bins, with the migration into the
95\%-exclusion bin now from approximately 15\% (single) to over 50\% (combined)
in both the expected and observed cases. Notably the wino scenario has a larger
fraction of expected single-SR points at $p$-values $> 0.3$, meaning there is a
larger population of points at moderate and high-$p$ able to be improved upon by
SR combination.

\begin{figure}
    \centering
    \subfloat[Expected $P(\mathrm{combined}|\mathrm{single})$]{{\includegraphics[width=0.475\textwidth]{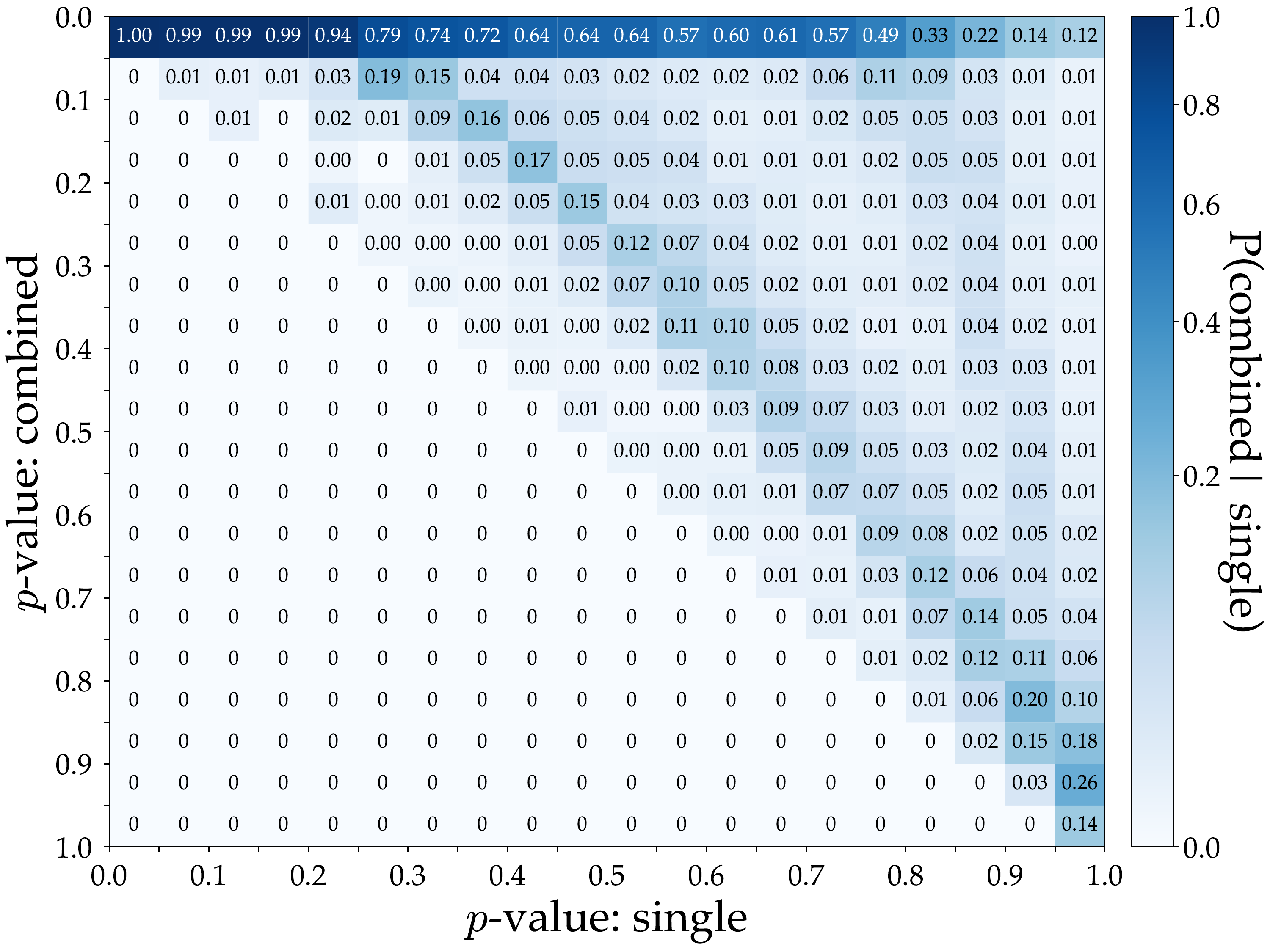} }}
    \subfloat[Expected $P(\mathrm{single}|\mathrm{combined})$]{{\includegraphics[width=0.475\textwidth]{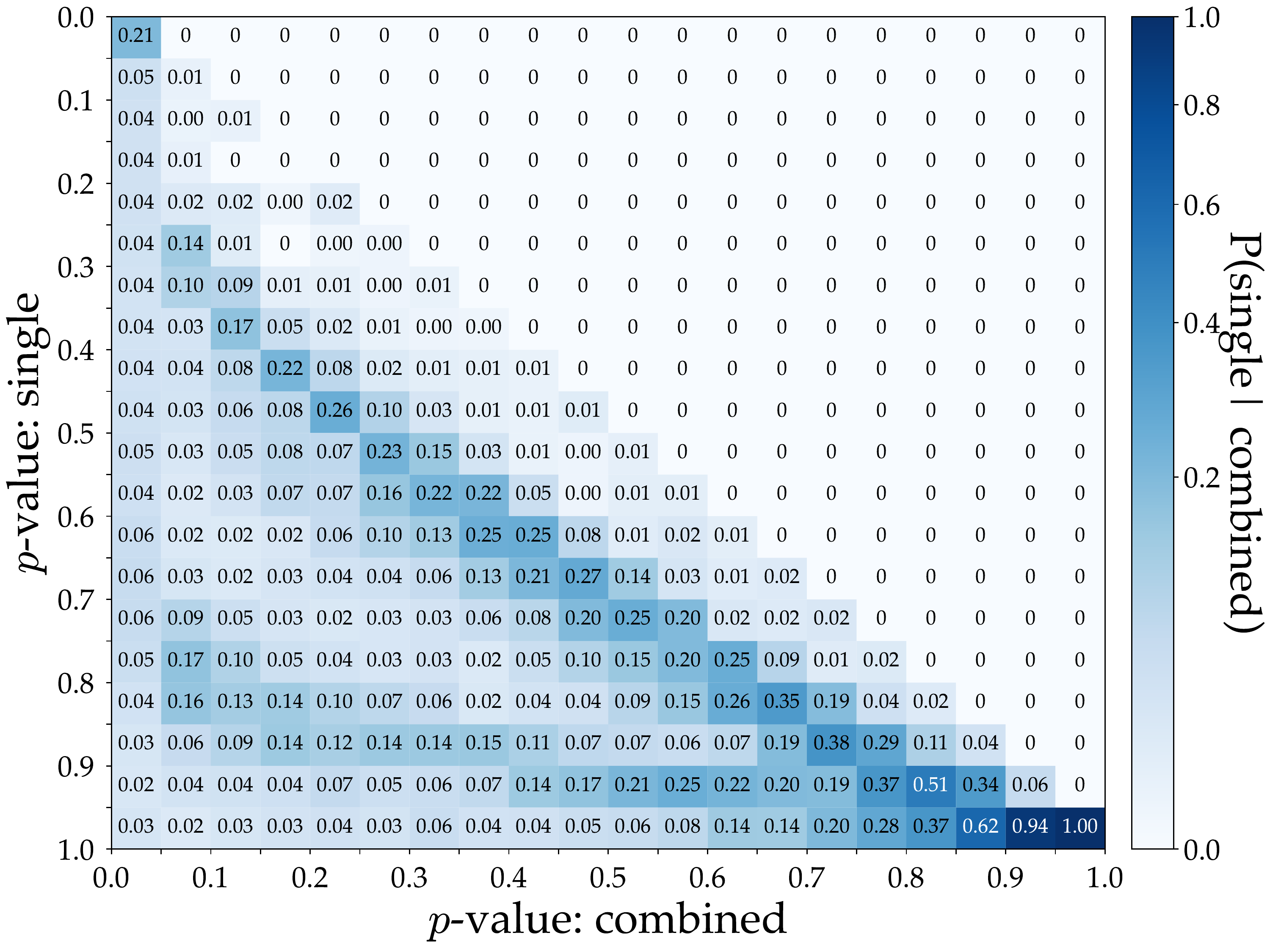} }}\\
    \subfloat[Observed $P(\mathrm{combined}|\mathrm{single})$]{{\includegraphics[width=0.475\textwidth]{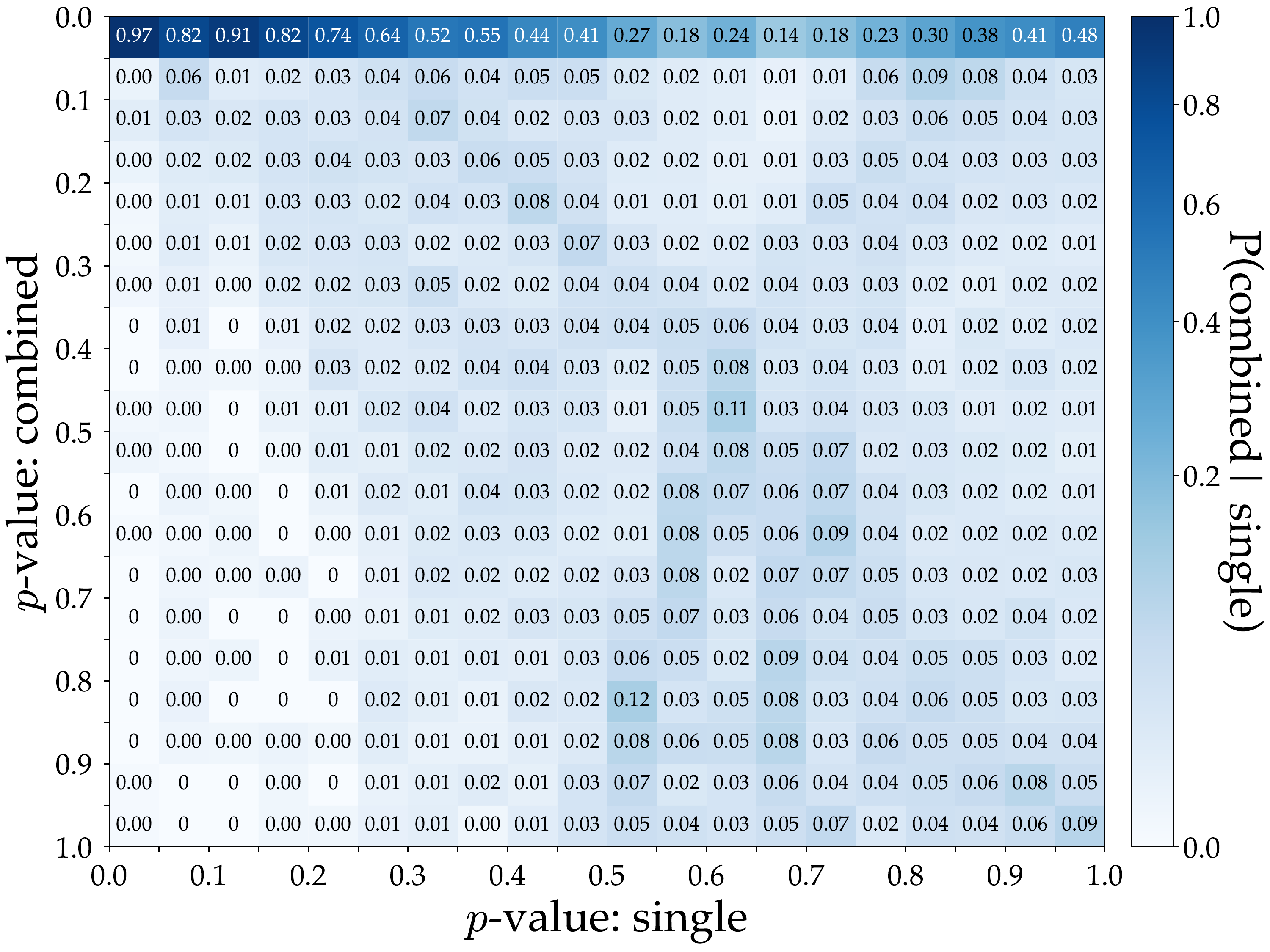} }}
    \subfloat[Observed $P(\mathrm{single}|\mathrm{combined})$]{{\includegraphics[width=0.475\textwidth]{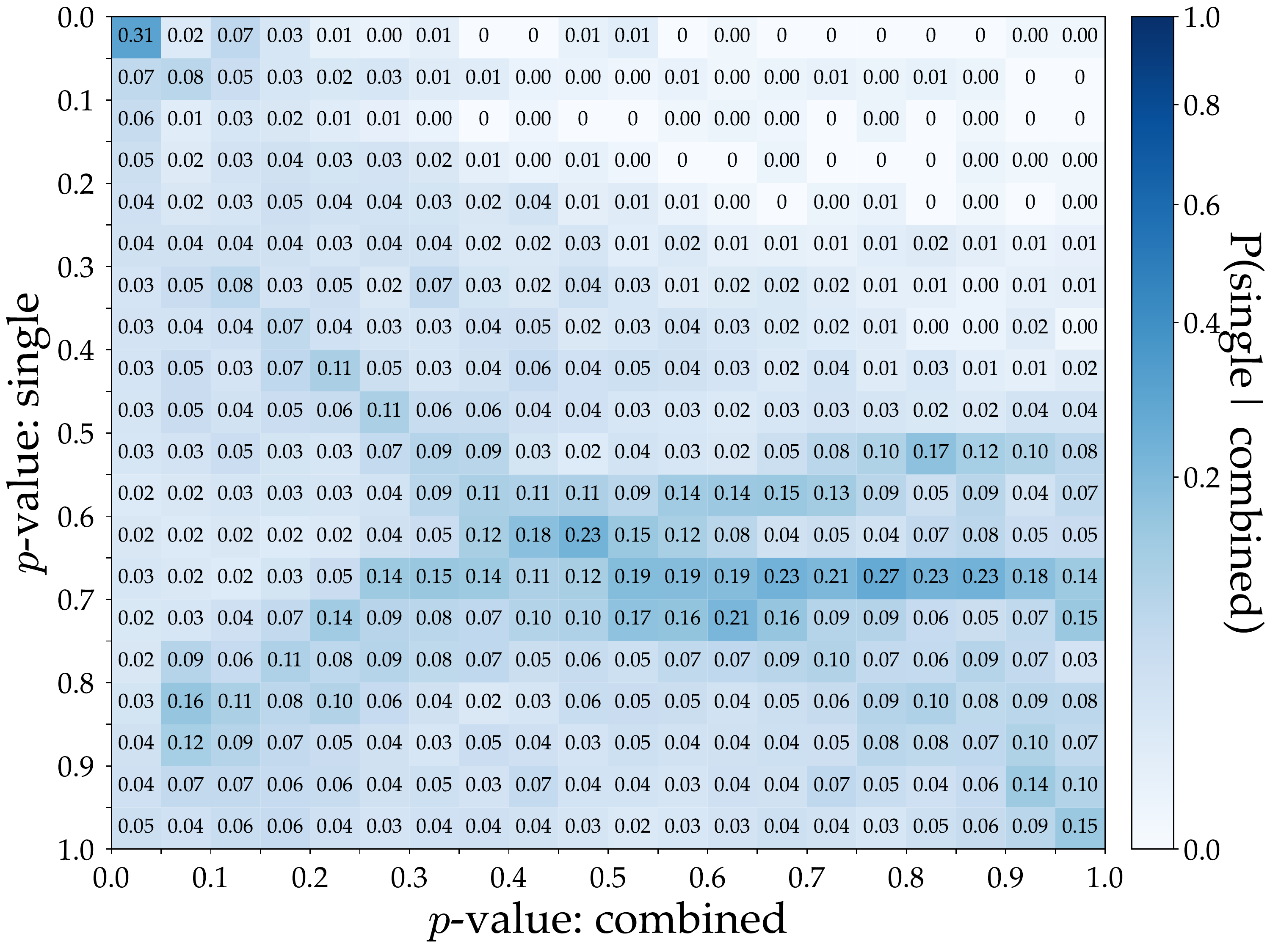} }}

    \caption{Transition matrices showing the pMSSM-19 wino results. The matrices
      describe the probability of a model point moving from one $p$-value bin to
      another, by use of the SR-combination scheme rather than the conservative
      single-best-SR strategy adopted by current recasting tools. The subfigure
      columns split the transition behaviours by the transition to combined
      $p$-value distributions given single-SR performance, and the single-SR
      origins of each combined-SR $p$-value range. The top and bottom subfigure
      rows show the expected and observed transitions respectively.
      }
    \label{fig: WINO Results}
\end{figure}

Considering the transition matrices shown in Figure~\ref{fig: WINO Results}, we
see similar trends to the bino case. The $P(\mathrm{combined} |
\mathrm{single})$ plots show the clear shift into the exclusion bin identified
in the 1D histogram, and the observed plots again contain a degree of negative
transitions, although not enough for the overall population of higher-$p$ bins
to increase.

The expected transitions into the 95\%-excluded $p$-value bin extend less far
along the single-SR $p$-value spectrum than in the bino-LSP case, with only 12\%
of the least-excluded single-SR points (those in the single-SR $p > 0.95$ bin)
expected to transition into the combined-SR exclusion bin. In practice, seen in
the observed-yield plots, over-fluctuations in SR yields led to greater
exclusion than expected for poorly constrained single-SR model points, with
nearly 50\% of the least excluded being eliminated in combination.


The filament structures in both the expected and observed sets of
transition plots are more prominent than in the wino case, allowing
identification of their origin. One, the shallower lower line in subfigure (b),
can be identified with the single-SR peak structures at $\sim \! 0.95$ and $\sim
\!  0.65$ for expected and observed respectively; the other is the main trend of
migration, showing that the dominant contribution to a given combined $p$-value
bin comes from a single-SR $p$-value bin 0.2 units higher.  These transition
structures again highlight potential for further improvements in model-point
exclusion fraction upon availability of more SRs.


\begin{figure}
    \centering
    \subfloat[Initial index (bino)]{{\includegraphics[width=0.475\textwidth]{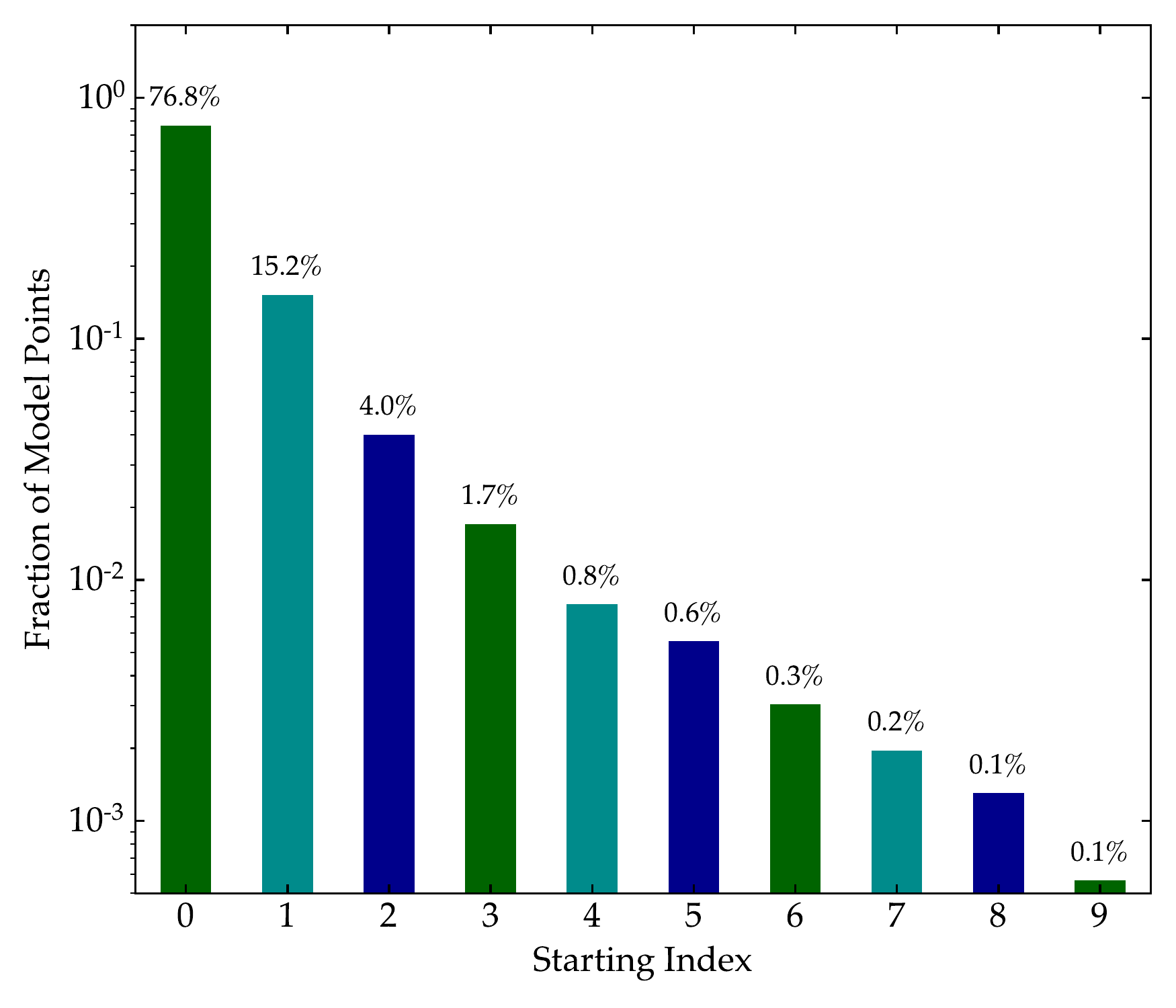} }}
    \subfloat[Initial index (wino)]{{\includegraphics[width=0.475\textwidth]{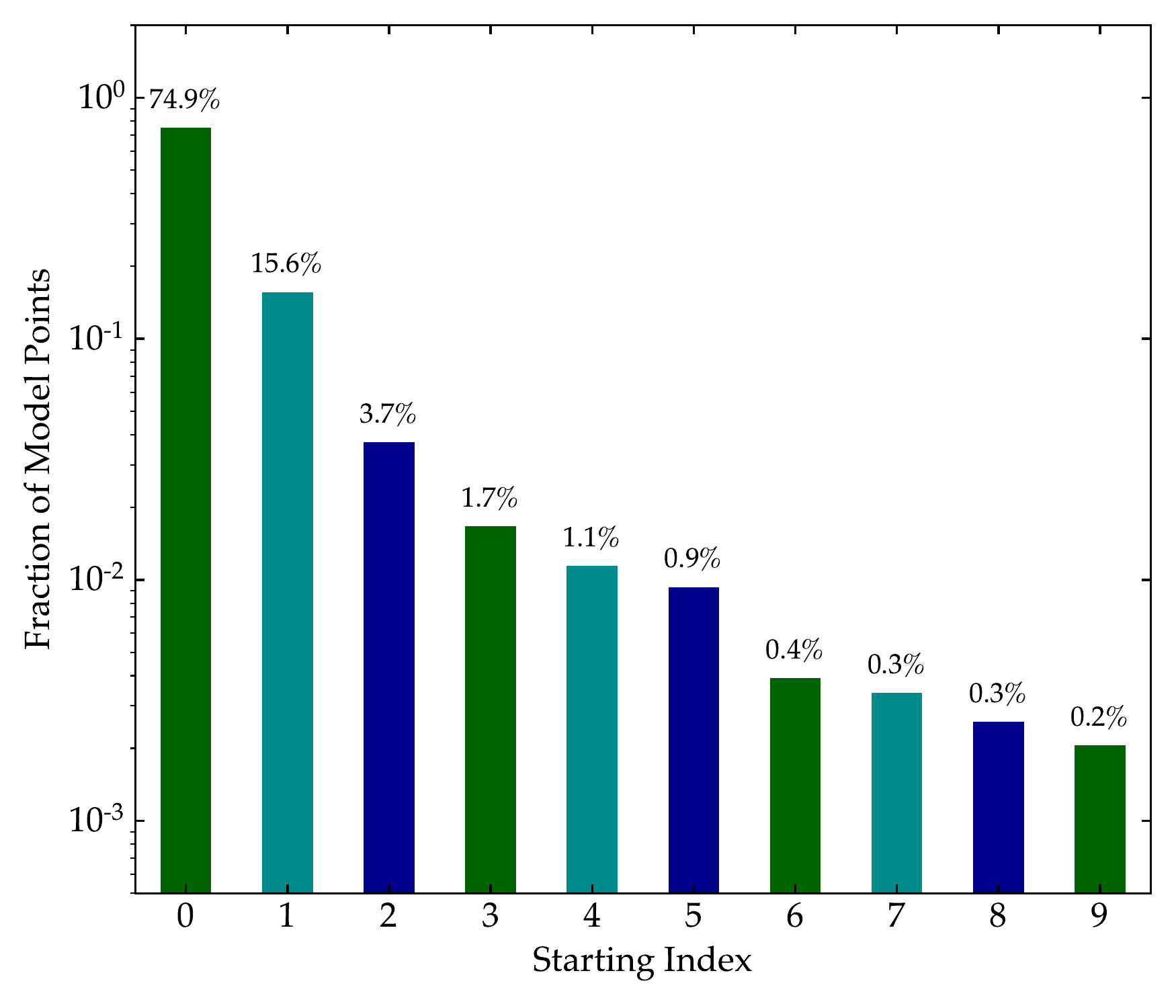} }}
    \hspace{0mm}
    \subfloat[Number of \SRs in combination (bino)]{{\includegraphics[width=0.475\textwidth]{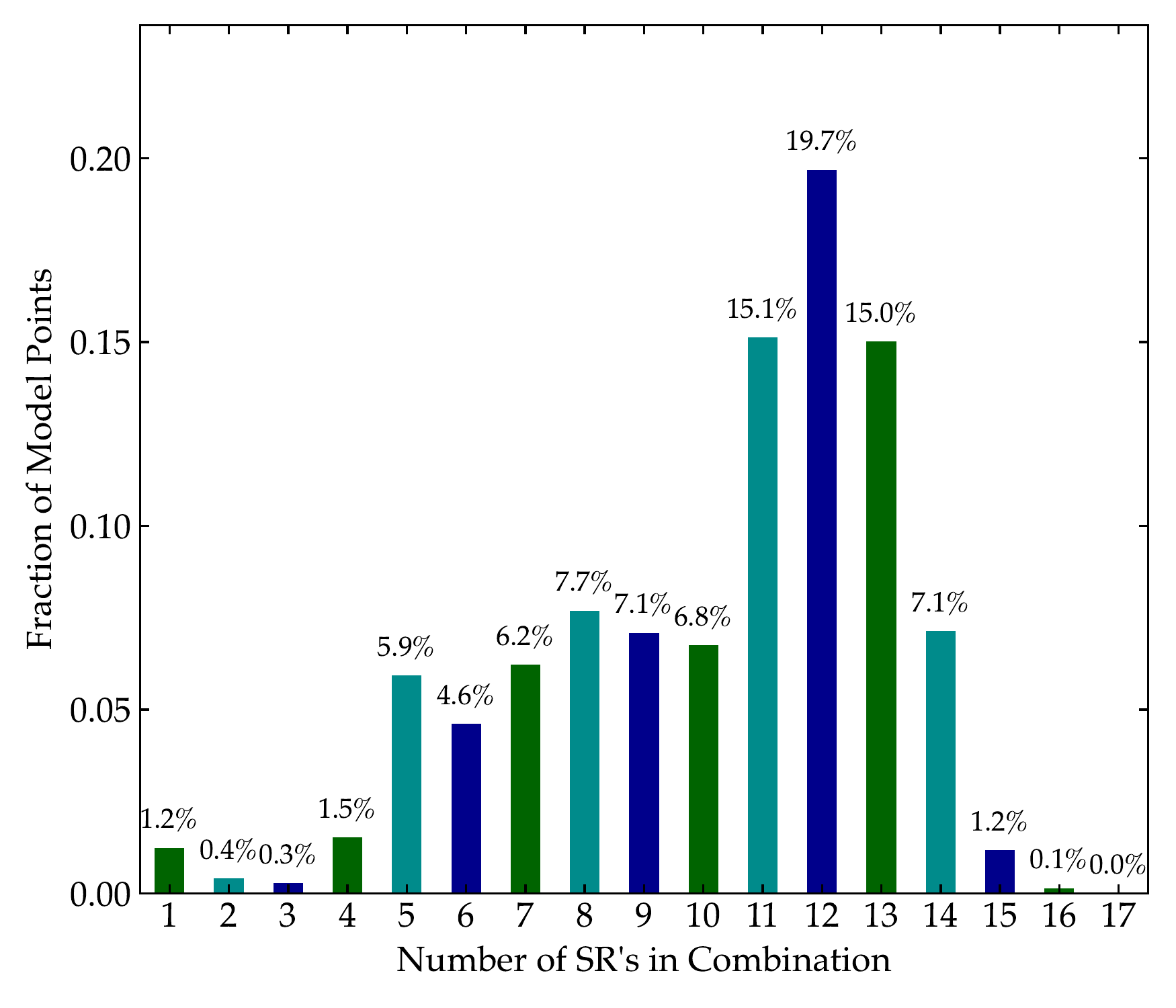} }}
    \subfloat[Number of \SRs in combination (wino)]{{\includegraphics[width=0.475\textwidth]{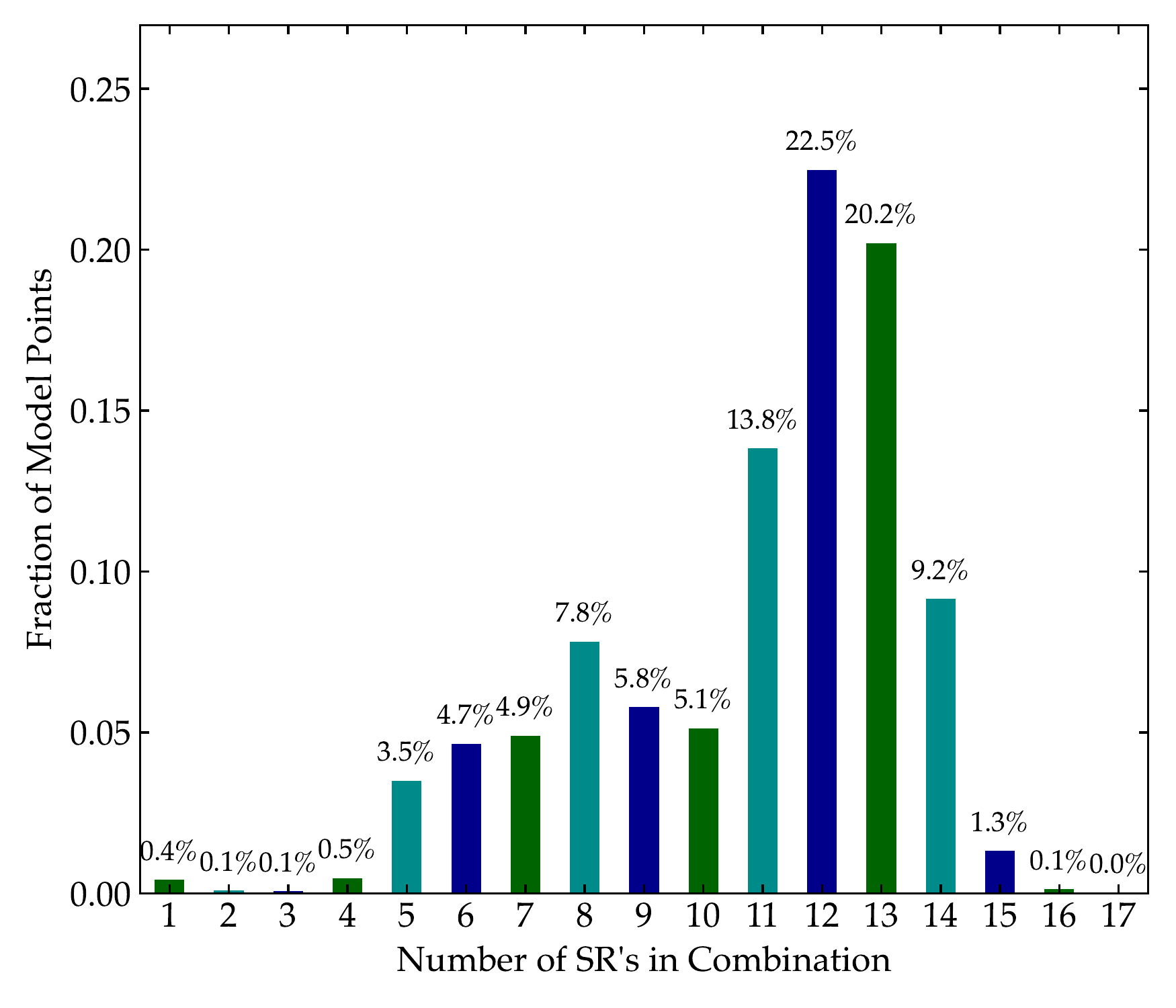} }}

    \caption{Fractional distributions of starting-SR~(a, b) and number of
      SRs~(c, d) in each combination. Data is taken from the optimum
      SR-combination found for each model-point for both the pMSSM-19 bino and
      wino reinterpretations. As mentioned in Section~\ref{sec:srcomb} the
      combinations are constructed from an differently ordered set of SRs for
      each point in the model space, so the identity of the zeroth SR is free to
      change from point to point.}
    \label{fig: pMSSM combinations}
\end{figure}

As in the simplified-model interpretation of Section~\ref{sec:res:t1tttt}, we
can examine the performance of the WHDFS SR-combination algorithm in both the
bino- and wino-LSP pMSSM-19 reinterpretations. The distributions of initial SR
indices are shown in the upper row of Figure~\ref{fig: pMSSM combinations}, with
the same bias toward low starting indices as in Figure~\ref{fig: T1 review}.


The lower two plots of Figure~\ref{fig: pMSSM combinations} show the
distribution of the number of SRs per combination for (c)~the bino and (d)~the
wino reinterpretations. The rapid fall-off of this distribution is a direct
consequence of the hereditary condition which reduces the number of available
SRs with each iteration of the path-building. Thus, there is a critical point
beyond which the cumulative drop-off of available SRs becomes statistically
evident in the mean number of SRs. For the pMSSM-19 bino and wino analyses, this
critical point occurs around 11~SRs.



\subsection{$t$-channel dark-matter}
\label{sec:res:tcdm}

As a final illustration of the strength of our approach, we consider in this
section one of the $t$-channel dark matter models explored in
Refs.~\cite{Arina:2020udz,Arina:2020tuw}. Here, the Standard Model is
extended by one fermionic dark-matter candidate~$\chi$ and one scalar mediator
state~$Y$, which interact with the right-handed up-quark. The model's Lagrangian
reads
\begin{equation}
  \mathcal{L} = \mathcal{L}_\mathrm{SM}  + {\cal L}_\mathrm{kin} +  \Big[ y  \big( \chi u_R \big)\ Y^\dag + \mathrm{H.c.} \Big] \, ,
\end{equation}
where $\mathcal{L}_\mathrm{SM}$ is the Standard-Model Lagrangian,
$\mathcal{L}_\mathrm{kin}$ contains kinetic and mass terms for all new states,
and $y$ dictates the strength of the interaction between the mediator, the dark
matter and the up-quark. In such a model, the \emph{full} new-physics signal
contains three contributions, namely
\begin{enumerate}
\item direct dark-matter production in association with one hard jet originating
  from initial-state radiation ($pp\to \chi\chi j$);
\item on-shell mediator-pair production followed by mediator decays into dark
  matter and jets ($pp\to YY^\ast\to \chi j \chi j$);
\item and the associated production of a mediator (that then decays into a $\chi
  j$ system) and a dark-matter state ($pp\to \chi Y \to \chi (\chi j)$).
\end{enumerate}

Such a signal can be searched for through analyses targetting the production of
multiple jets in association with missing transverse-energy, each component of
the signal yielding a different jet-multiplicity spectrum and different jet
properties. We therefore focus on the reinterpretation of the results of the
ATLAS-SUSY-2015-06~\cite{ATLAS:2016dwk},
ATLAS-SUSY-2016-07~\cite{ATLAS:2017mjy}, CMS-SUS-16-033~\cite{CMS:2017abv} and
CMS-SUS-19-006~\cite{CMS:2019zmd} analyses to investigate which mediator and
dark-matter mass configurations are allowed by data, for a new-physics coupling
set to $y=1$. All analyses considered are integrated in the \MA Public Analysis
Database~\cite{Dumont:2014tja}, the recast codes and their detailed validation
notes being available from
Refs.~\cite{DVN/MHPXX4_2021,DVN/I5IFG8_2021,DVN/GBDC91_2021,DVN/4DEJQM_2020,Mrowietz:2020ztq}
and on the database webpage.\footnote{See
\url{http://madanalysis.irmp.ucl.ac.be/wiki/PublicAnalysisDatabase}.}

To estimate the individual exclusion limits originating from each analysis, we
used \amc~v2.6.5~\cite{Alwall:2014bza} to generate hard-scattering events at
leading-order (LO) accuracy. We grouped the three different contributions to the
signal in two sets according to the parton-level jet multiplicity. A first
matrix-element describes the production of a pair of dark-matter states with a
single hard jet ($pp \to \chi\chi j$), and a second one concerns mediator
pair-production and decay ($pp\to YY^\ast\to \chi j \chi j$). The associated
production of a dark-matter particle with a mediator is hence included in the
first subprocess, as it yields the same final-state ($pp\to \chi Y \to \chi
(\chi j)$ with the intermediate mediator $Y$ being on-shell). These two
matrix-elements were convolved with the NNPDF~2.3~LO~\cite{Ball:2013hta} set of
parton distribution functions, and we generated \num{200000} signal events per
model-point to limit statistical uncertainties. Hadronisation and parton
showering were handled with \pythiaE~v8.240~\cite{Sjostrand:2014zea} and the
simulation of the CMS and ATLAS detector responses was approximated with
\delphes~\cite{deFavereau:2013fsa}, using the custom detector-parameterisation
provided with each recast code.

\begin{figure}
    \centering
    \includegraphics[width=0.49\textwidth]{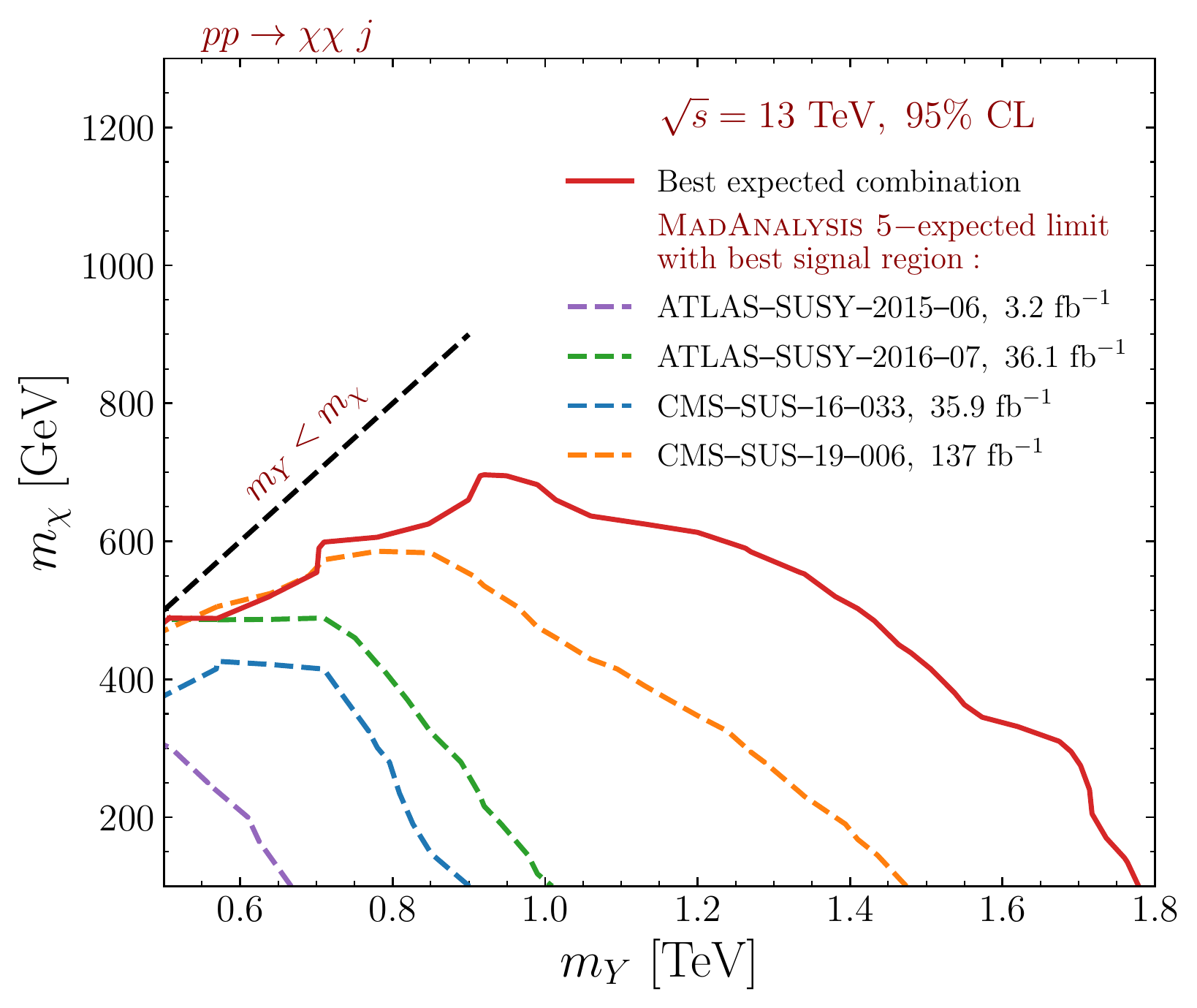}\hfill
    \includegraphics[width=0.49\textwidth]{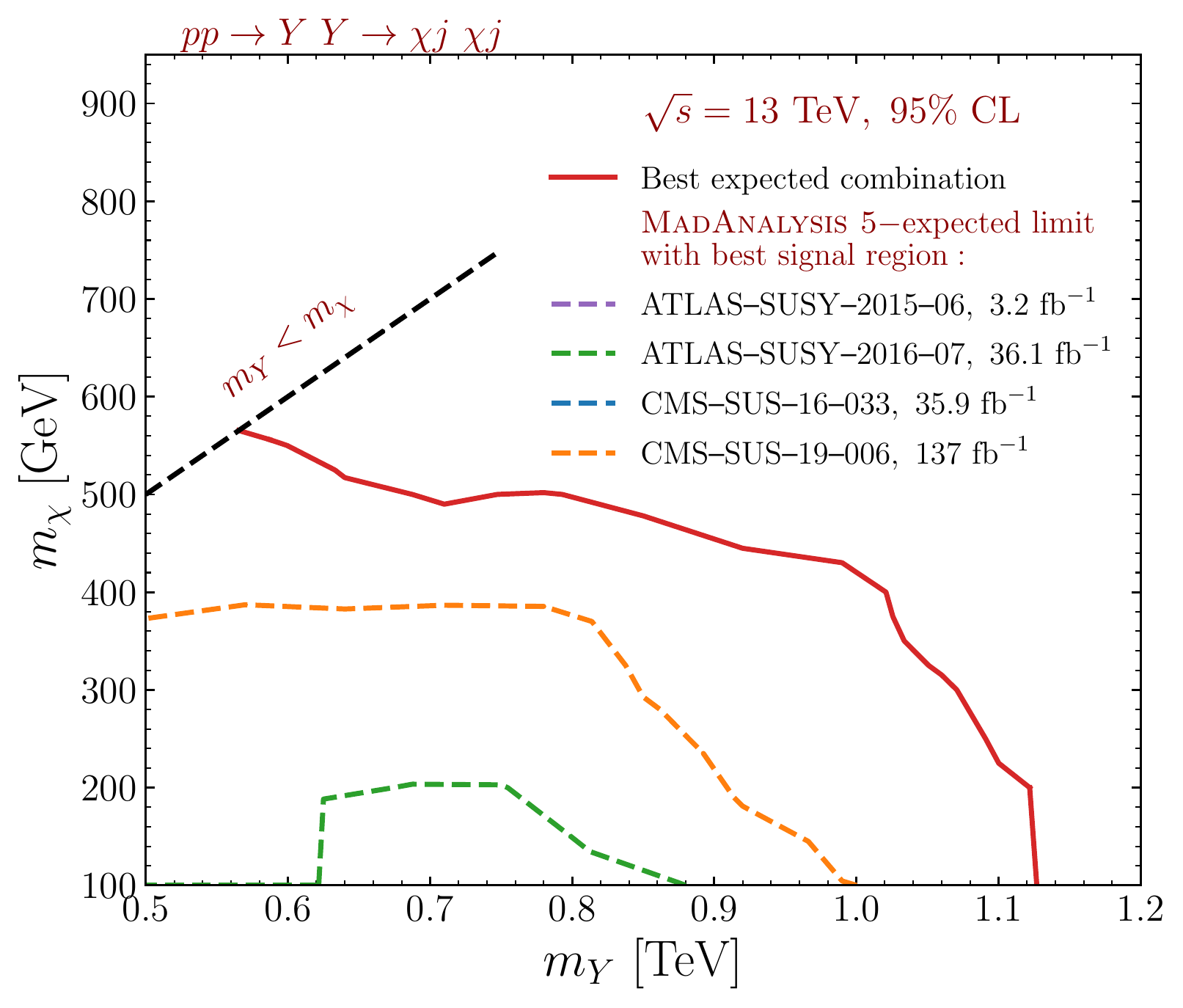}\\
    \includegraphics[width=0.49\textwidth]{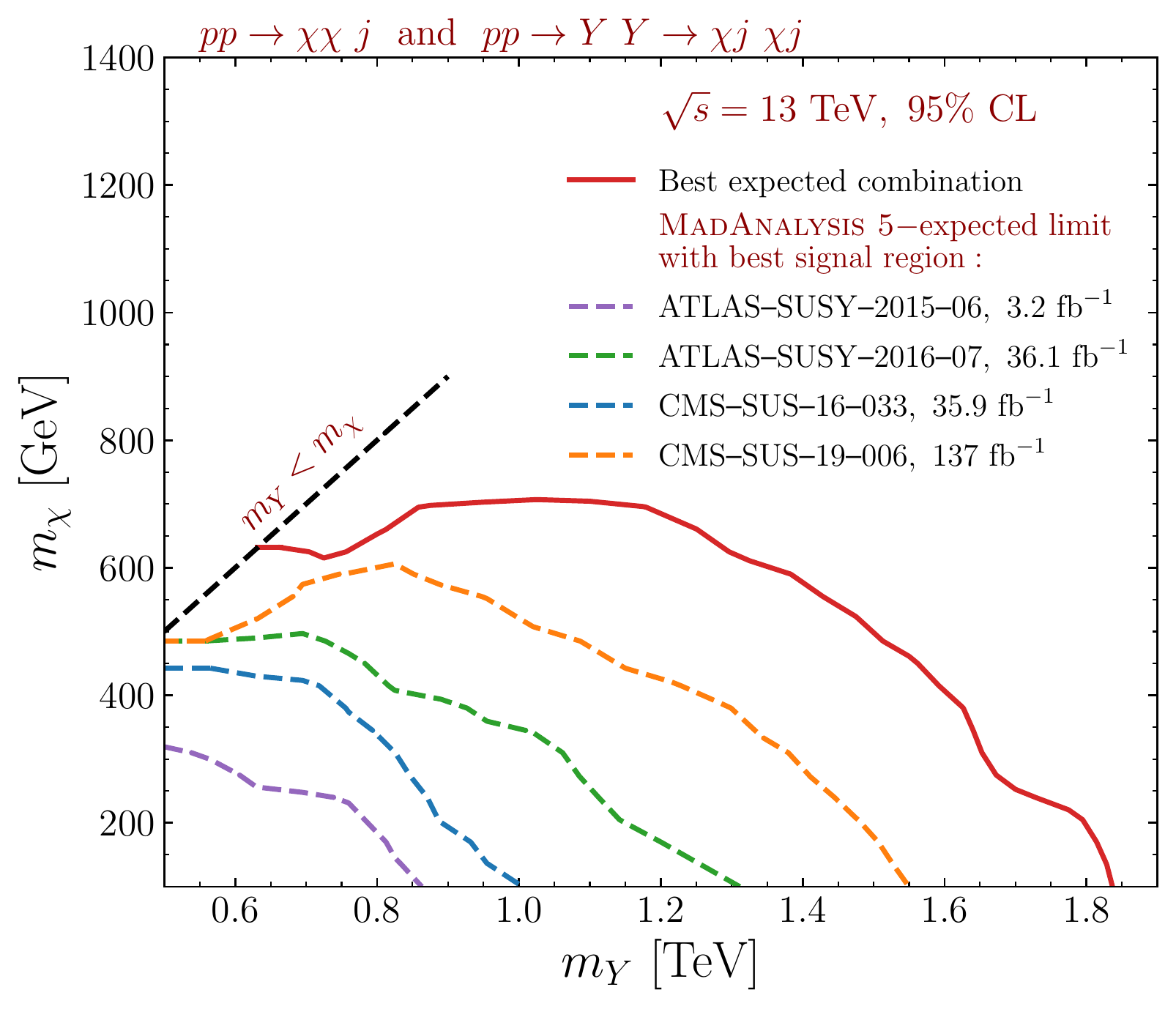}
    \caption{95\% confidence level exclusions for the studied $t$-channel
      simplified model of dark matter. We explore separately the two components
      of the new-physics signal respectively arising from the processes $pp\to
      \chi\chi j$ (upper left) and $pp\to YY^\ast\to \chi j \chi j$ (upper
      right), as well as their sum (lower panel). We display exclusion limits
      originating from the individual recast analyses, \ie~for the
      ATLAS-SUSY-2015-06 (dashed purple line), ATLAS-SUSY-2016-07 (dashed green
      line), CMS-SUS-16-033 (dashed blue line) and CMS-SUS-19-006 (dashed orange
      line) analyses, as well as those derived from their combination through
      the method proposed in this paper (solid red
      line).}\label{fig:ma5_results}
\end{figure}

The results are displayed in the $(m_Y, m_\chi)$ plane in
Figure~\ref{fig:ma5_results}, for a mediator mass $m_Y \in [0.5, 1.8]~\si{\TeV}$
and a dark-matter mass $m_\chi\in [0.1, 1]~\si{\TeV}$. We separately show
exclusions extracted from single-jet events only ($pp\to \chi\chi j$, upper left
panel) and from dijet events only ($pp\to YY^\ast\to \chi j \chi j$, upper right
panel), as well as the combined limits derived by considering the \emph{full}
new-physics signal (lower panel). In these three figures, the dashed lines
represent the individual limits obtained by the reinterpretation of the results
of the ATLAS-SUSY-2015-06 (purple), ATLAS-SUSY-2016-07 (green), CMS-SUS-16-033
(blue) and CMS-SUS-19-006 (orange) analyses, which respectively probe integrated
luminosities of \SI{3.2}{\ifb}, \SI{36.1}{\ifb}, \SI{35.9}{\ifb}, and
\SI{137}{\ifb}. These limits were obtained by conservatively considering the
signal region of a given analysis giving rise to the best expected exclusion,
for a given benchmark point.

Unsurprisingly, the analysis making use of the largest amount of data,
CMS-SUS-19-006, leads to the strongest individual exclusion.
Considering only the ``single-jet'' component of the signal
(Figure~\ref{fig:ma5_results}, upper left panel), mediator masses up to
\SI{1.5}{\TeV} are excluded by the CMS-SUS-19-006 analysis for small dark-matter
masses $m_\chi$. By comparison, the ATLAS-SUSY-2016-07 and CMS-SUS-16-033
analyses which analysed only one third of the Run~2 integrated luminosity, are
only sensitive to mediator masses smaller than \SIrange{900}{1000}{\GeV} for the
same $m_\chi$ assumptions. Similarly, scenarios with more compressed spectra,
which are intrinsically harder to probe as they correspond to the production of
softer final-state objects, see better coverage from the most recent analysis
than from the two partial Run~2 analyses. By contrast to all of these, the
early- Run~2 ATLAS-SUSY-2015-06 analysis barely reaches exclusion for
new-physics masses at \SI{500}{\GeV}.

Similar conclusions hold for the ``dijet'' component of the signal
(Figure~\ref{fig:ma5_results}, upper right panel). The most recent
CMS-SUS-19-006 analysis is sensitive to both larger new-physics masses and more
compressed spectra than the partial Run~2 analyses --- the latter are in this
case barely sensitive to any signal. The sensitivity is found to be
significantly milder than for the single-jet signal component, due to the
phase-space suppression associated with the production of two heavy mediators.
Consequently, it turns out that for the range of masses to which Run~2 of the
LHC is sensitive, the limits obtained for the full new-physics scenario
(Figure~\ref{fig:ma5_results}, lower panel) are almost identical to these
obtained in the single-jet scenario. For light dark matter, we observe an
improvement of about \SI{100}{\GeV}; this demonstrates that even when it yields
mild effects, use of the full new-physics signal is always better than an
approximate modelling that does not include all relevant subprocesses.

Figure~\ref{fig:ma5_results} also shows the impact of combining the four
analyses, first for the individual components of the new-physics signal (upper
panel), and next after combining them (lower panel). This was enabled by
determining and using the overlaps between the various analyses' signal regions
cf.~Sections~\ref{sec:correst} and~\ref{sec:srcomb}.
Despite the fact that the analyses targeted similar topologies (multijet and
missing energy), some of their signal regions proved uncorrelated enough for
combination to be performed. This ilustrates the advantage of an objective and
quantified measure of acceptance overlap in place of an informal guesstimate
of orthogonality. The allowed level of combination allows for an increase in parameter-space
coverage,
as displayed by the solid red contours in Figure~\ref{fig:ma5_results}. We
observe a substantial improvement of the limits through WHDFS SR-combination,
both for split-mass spectra (light dark-matter and heavy mediator) and
compressed spectra.  For $m_\chi \approx \SI{100}{\GeV}$, mediator masses
ranging up to \SI{1.9}{\TeV} are reachable, whereas scenarios with a dark-matter
mass $m_\chi < \SI{600}{\GeV}$ get excluded for $m_Y < \SI{1.2}{\TeV}$.
\section{Conclusions and outlook}

In this paper we have argued that maximising the BSM-search power of the LHC and
other colliders, in the light of Run~2's significant list of null-result direct
searches, necessitates combination of analyses for sensitivity to more subtle
dispersed-signal models than have so far been considered.

But combination cannot be performed naively, due to overlaps in analyses' event
acceptance. Short of many-year, top-down coordination within experimental
collaborations to forbid phase-space overlaps (with implicit prioritisation of
some analyses over others) or public lists of which collider-event numbers
entered which signal regions across all published analyses, a \emph{post hoc}
method is needed to estimate the extent of such overlaps. In this paper we have
presented the \taco method for this estimation, using the \smodels and \MA
analysis databases to guide simulated-event population of all recastable signal
regions. A new augmentation to the \MA analysis machinery enables the overlap
coefficients between pairs of signal regions to then be computed via Poisson
bootstrapping.

We have also shown how this information can be used in a scalable way to obtain
the expected optimally model-excluding, non-overlapping subset of signal regions
for a given BSM model or model-point. The combinatorically hard problem of
evaluating all allowed subsets of $\mathcal{O}(400)$ signal regions (at present,
and guaranteed to grow) is mapped to construction of directed acyclic graphs
representing SR combinations.
This construction is made tractable and even rapid by use of a binarised form of
the SR-overlap matrix to efficiently exclude sequences of partially overlapping
graphs, and by ordering SRs in the graph construction according to their
expected log likelihood-ratios. The expected best-sensitivity combination can
then be efficiently identified by a weighted hereditary depth-first search
(WHDFS) algorithm, in direct analogy to a weighted longest-path problem.

Code for both the overlap estimation and SR-combination aspects of this paper is
publicly available from the \url{https://gitlab.com/t-a-c-o/taco_code} online
repository.

Computation of such SR-subsets using the \taco WHDFS algorithm is a practical
alternative to direct use of the overlap matrix to compute a correlated $\chi^2$
or other measure across all signal regions, given the latter approach's high
risk of numerical instabilities in covariance inversion, and the huge
computational cost of simultaneously likelihood-profiling over a full set of
SRs. This graph-based approach hence has potential not just as a route to
composite likelihoods in reinterpretation, but also as a dimension-reduction
technique for visualising and interpreting how dominant categories of analyses
and signal-regions evolve through model spaces. 

We have tested the \taco overlap-estimation and optimal-subset computations
against several BSM models of increasing complexity: a SUSY simplified model,
ATLAS' \SI{8}{\TeV} scan of pMSSM-19 points re-evaluated on \SI{13}{\TeV}
measurements, and a $t$-channel dark-matter model. In all cases we see the
algorithmic combination of SRs providing a significant increase in experimental
limit-setting reach, typically $\mathcal{O}(100)\,\si{\GeV}$ in the mass
parameters of both the simple and complex BSM models considered with currently
available reinterpretation analyses. The study of transition matrices for the SR
combination indicates that the gains obtained are not just marginal --- moving
already near-exclusion model points over the line --- but holistic, with
evidence that combination of up to ten relatively weak signal regions can create
a complementary strong limit.

This method hence demonstrates that \emph{post hoc} combination of BSM
direct-search data is possible and can be made computationally efficient, and
that pessimistic use of at most one signal-region from each event topology is no
longer necessary. As an efficient and empirical computational method, \taco is
scalable to hundreds of potentially overlapping analyses, beyond the capacity of
manual and fallible assessment of uncorrelated analysis sets. It is, of course,
imperfect. The assumption of effective orthogonality of finitely overlapping
$\rho_{ij} < T$ signal regions is key to efficient computation of SR subsets,
but a hybrid of subset-identification with correlated LLR evaluation on the
reduced set is easily appended to the procedure described here. Systematic
uncertainties are also missing from the current treatment, but --- at least on
the 
subset of uncertainties that can be evaluated by event
reweighting --- this is again not an intractable problem. We hope that this
method and toolkit will prove a useful target for how collider BSM-combinations
are designed and performed in the coming years, with submission of analysis
routines to the key reinterpretation-analysis frameworks and provision of
event-bootstrapping machinery beyond \MA.







\section*{Acknowledgements}

AB and JY thank the Royal Society for studentship funding under URF Enhancement
Grant RGF\textbackslash{}EA\textbackslash{}180252, and STFC for support under
Consolidated Grant ST/S000887/1. The research of HRG is funded by the Italian
PRIN grant 20172LNEEZ. The authors thank the Les Houches Physics at the
Terascale series for its 2019 edition during which this work originated, and we
look forward to the return of collider-physics to the academically productive
slopes of the Chamonix valley. Our further thanks to Sabine Kraml and others at
LPSC Grenoble for generously bringing the authors together to invigorate this
work, with support from the IN2P3 project ``Th\'eorie -- BSMGA''.


\clearpage
\renewcommand{\thesection}{\Alph{section}}
\setcounter{section}{0}
\appendix
\section{BSM-search overlap matrices}
\label{app:overlaps}

\begin{figure}[H]
    \centering
    \includegraphics[width=0.95\textwidth]{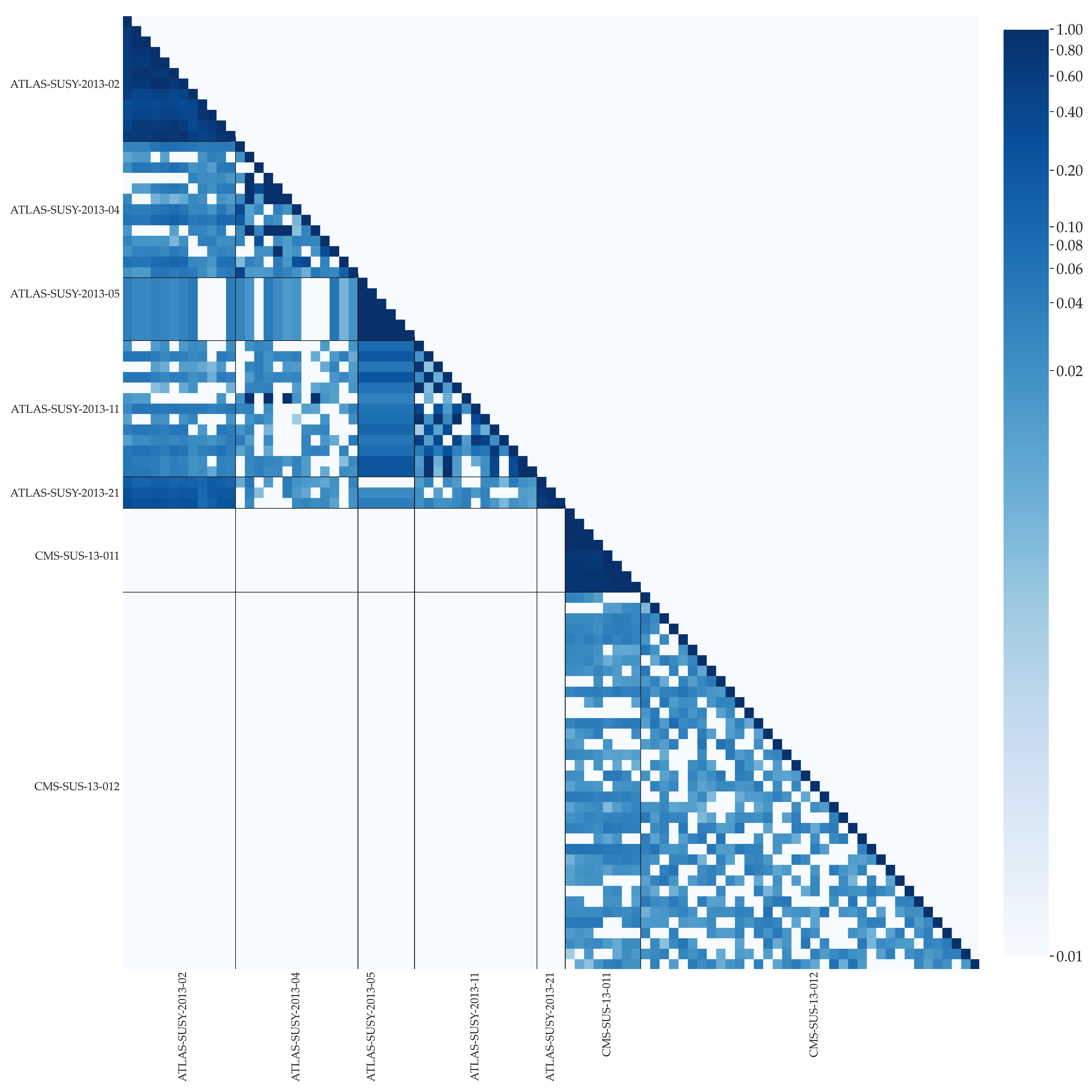}
    \caption{The overlap matrix $\rho_{ij}$ obtained from the TACO sampling procedure
      between all LHC BSM searches at \SI{8}{\TeV} commonly implemented in
      \smodels and \MA.  Non-overlap between ATLAS and CMS analyses is manually
      imposed, as the same proton-collisions could not be accepted by analyses
      from both experiments regardless of the MC overlaps, and similarly
      overlaps between 8~and \SI{13}{\TeV} analyses must be zero regardless of
      final-state acceptances. The set of SR pairs considered sufficiently
      independent in the analyses of Sections~\ref{sec:srcomb}
      and~\ref{sec:results}, with $T < 0.01$, are shown in white.}
    \label{fig:overlaps8}
\end{figure}

\begin{figure}
    \centering
    \includegraphics[width=0.95\textwidth]{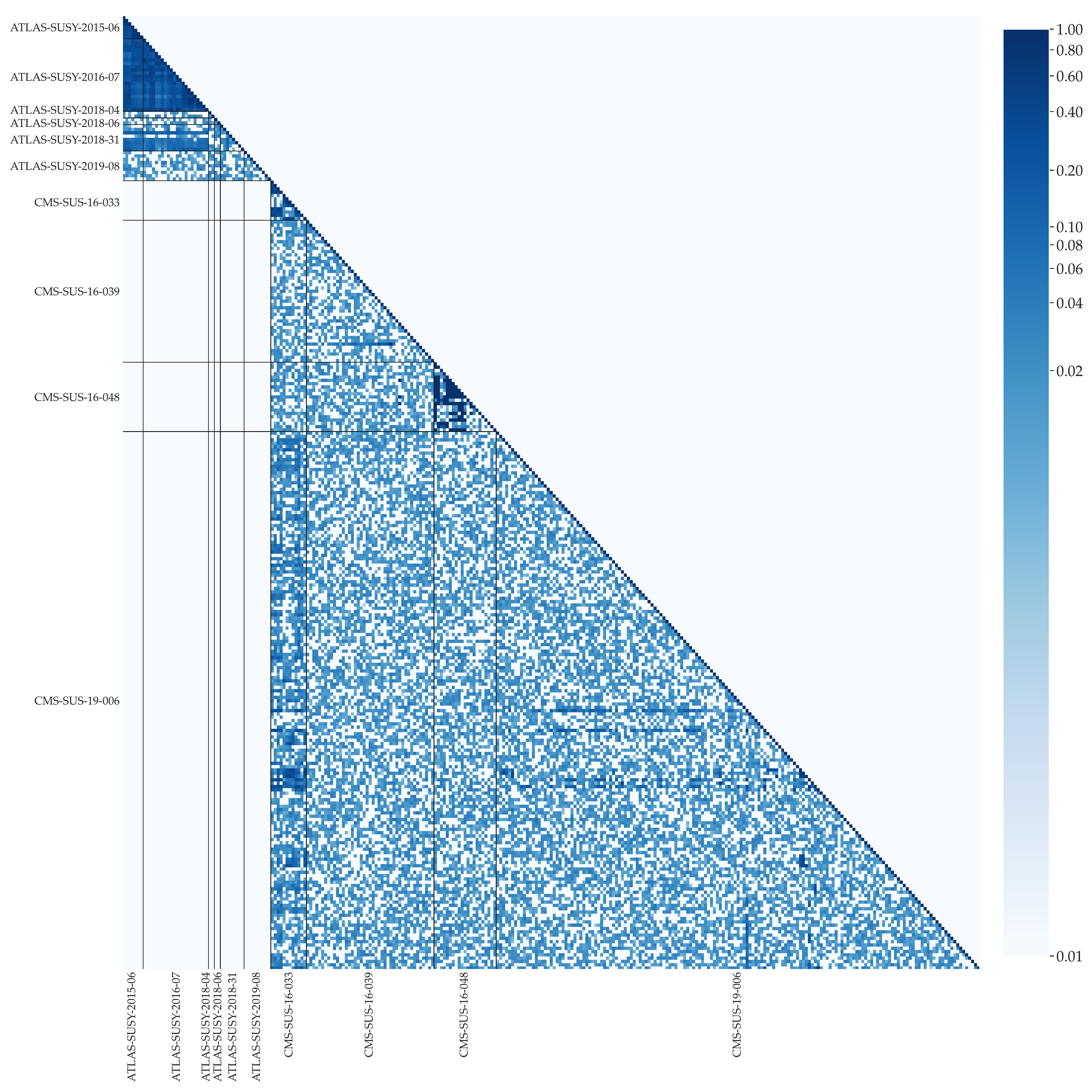}
    \caption{The overlap matrix $\rho_{ij}$ obtained from the TACO sampling procedure
      between all LHC BSM searches at \SI{13}{\TeV} commonly implemented in
      \smodels and \MA.  Non-overlap between ATLAS and CMS analyses is manually
      imposed, as the same proton-collisions could not be accepted by analyses
      from both experiments regardless of the MC overlaps, and similarly
      overlaps between 8~and \SI{13}{\TeV} analyses must be zero regardless of
      final-state acceptances. The set of SR pairs considered sufficiently
      independent in the analyses of Sections~\ref{sec:srcomb}
      and~\ref{sec:results}, with $T < 0.01$, are shown in white.}
    \label{fig:overlaps13}
\end{figure}

\clearpage

\section{HDFS algorithm}
\label{app:alg}
\label{app:hdfsalg}

The pseudocode shown in Algorithms~\ref{alg:DHAG} and~\ref{alg:WDHAG} are written in a Pythonic syntax as the code makes use of the generator -- donoted by the term \textit{Gen()} --
functionality which allows for efficient iteration ordering.
Aspects of the code are heavily influenced by the ``all simple paths'' method
from the Python \sw{NetworkX} package~\cite{SciPyProceedings_11}.

\begin{algorithm}
\caption{Hereditary Depth-First Search (HDFS)}\label{alg:cap}
\label{alg:DHAG}
\bigskip
\begin{algorithmic}

\Require source = i
\State Target = n
\State Cutoff = $n-1$
\State Visited = $[i]$
\State Stack = $[Gen(A_i)]$
\State $\mathcal{S}$ =  $[A_{i}]$
\While{Stack is not empty}
    \State Children = last element of stack
    \State c = Next element in Children  \textbf{or} None \textbf{if} Empty
    \If{c is None}
        \State Drop last element of Stack
        \State Drop last element of  $\mathcal{S}$
        \State Drop last element of Visited
    \ElsIf{length(Visited) $<$ cutoff}
        \If{c = Target}
            \State \textbf{Yield}: Visited + [c]
        \EndIf
        \State Visited += [c]
        \If{target not in Visited}
            \State $\mathcal{S}_c$ = $A_c \cap S_{c-1}$
            \State Stack += [Gen(j: \textbf{for} index in $A_c$ \textbf{if} index $\in$ $\mathcal{S}_c$)]
        \ElsIf{}
            \State Drop last element of Visited
        \EndIf
    \ElsIf{length(Visited) $=$ cutoff}
        \State Drop last element of Stack
        \State Drop last element of  $\mathcal{S}$
        \State Drop last element of Visited
        \State \textbf{Yield}: Visited + [Target]
    \EndIf
\EndWhile
\end{algorithmic}
\end{algorithm}

\clearpage

\section{WHDFS algorithm}
\label{app:whdfsalg}

The pseudocode shown in Algorithm~\ref{alg:WDHAG} is a modification of Algorithm~\ref{alg:DHAG}. WHDFS uses the edge weights to calculate an upper limit of total weight available at each step in the path. This modification eliminated the need to explore all allowed paths instead limiting the combinations to those that have the greatest potential.

\begin{algorithm}[H]
\caption{Weighted Hereditary Depth-First Search (WHDFS)}\label{alg:cap1}
\label{alg:WDHAG}
\bigskip
\begin{algorithmic}

\Require source = i
\Require maximum weight
\State Best Path = []
\State Target = n
\State Cutoff = $n-1$
\State Visited = $[i]$
\State Stack = $[Gen(A_i)]$
\State $\mathcal{S}$ = $[A_{i}]$
\While{Stack is not empty}
    \State Children = last element of stack
    \State c = Next element in Children  \textbf{or} None \textbf{if} Empty

    \If{c is None}
        \State Drop last element of Stack
        \State Drop last element of  $\mathcal{S}$
        \State Drop last element of Visited
    \ElsIf{length(Visited) $<$ cutoff}
        \If{Target in Visited}
            \State continue
        \EndIf

        \State $\mathcal{S}_c$ = $A_c \cap S_{c-1}$
        \State Visited += [c]
        \State Current Weight = weight function (Visited)
        \State Available Weight = weight function ($\mathcal{S}_c$)
        \If{c = Target \& Current Weight $>$ Max weight}
            \State Max weight = Current Weight
            \State Best Path = Visited
        \EndIf
        \If{(Current Weight + Remaining Weight) $>$ Max weight}
            \State Stack += [Gen(j: \textbf{for} index in $A_c$ \textbf{if} index $\in$ $\mathcal{S}_c$)]
        \ElsIf{}
            \State Drop last element of Visited
        \EndIf
    \ElsIf{length(Visited) $=$ cutoff}
        \State Drop last element of Stack
        \State Drop last element of  $\mathcal{S}$
        \State Drop last element of Visited
    \EndIf
\EndWhile
\State \Return Best Path
\end{algorithmic}
\end{algorithm}

\clearpage
\bibliography{taco21}

\end{document}